\documentclass[twocolumn,aps,floatfix,superscriptaddress]{revtex4-1}
\usepackage{amsmath,amssymb,eucal,graphicx,float,epstopdf,color,fancyhdr}

\begin{document}

\title{Variational calculation of transport coefficients in diffusive lattice gases}

\def\shorttitle{Variational calculation of transport coefficients}

\author{Chikashi Arita}
\affiliation{Theoretische Physik, Universit\"at
 des Saarlandes, 66041 Saarbr\"ucken, Germany}
\author{P. L. Krapivsky} 
\affiliation{Department of Physics, Boston University, Boston, MA 02215, USA}
\affiliation{Institut de Physique Th\'eorique, IPhT, CEA Saclay
and URA 2306, CNRS, 91191 Gif-sur-Yvette cedex, France}
\author{Kirone Mallick} 
\affiliation{Institut de Physique Th\'eorique, IPhT, CEA Saclay
and URA 2306, CNRS, 91191 Gif-sur-Yvette cedex, France}

\def\shortauthor{C. Arita, P. L. Krapivsky, K. Mallick}

\pagestyle{fancy}
 \lhead{\shorttitle} \chead{} \rhead{\shortauthor} 
 \renewcommand{\headrulewidth}{0.5pt} 
 \renewcommand{\footrulewidth}{0.5pt} 
 \lfoot{} \cfoot{\thepage} \rfoot{} 
 
 \fancypagestyle{titlepage}{
 \renewcommand{\headrulewidth}{0pt} 
 \renewcommand{\footrulewidth}{0.5pt} 
 \lhead{} \chead{} \rhead{}
 \lfoot{} \cfoot{\thepage} \rfoot{} }

\begin{abstract}
A diffusive lattice gas is characterized by the diffusion coefficient depending only on the density. The Green-Kubo formula for diffusivity can be represented as a variational formula, but even when the equilibrium properties of a lattice gas are analytically known the diffusion coefficient can be computed only in the exceptional situation when the lattice gas is gradient. In the general case, minimization over an infinite-dimensional space is required. We propose an approximation scheme based on minimizing over finite-dimensional subspaces of functions. The procedure is demonstrated for one-dimensional generalized exclusion processes in which each site can accommodate at most two particles. Our analytical predictions provide upper bounds for the diffusivity that are very close to simulation results throughout the entire density range. We also analyze non-equilibrium density profiles for finite chains coupled to reservoirs. The predictions for the profiles are in excellent agreement with simulations. 
\end{abstract}

\maketitle 
\thispagestyle{titlepage}

\newcommand{\E}{{\mathbb E}}
\renewcommand{\P}{{\mathbb P}}

\section{Introduction} 
\label{sec:}

In systems composed of a huge number of interacting particles, a continuous macroscopic behavior emerges after space and time variables are suitably rescaled. In this {\it hydrodynamic limit}, the intrinsic granularity of the basic constituents is lost and the system is described by continuous fields (such as matter or charge densities, currents, magnetization, etc.), coupled by partial differential equations. The program of developing a mathematical theory of hydrodynamic limits for general systems was posed by Hilbert \cite{bib:Hilbert} in his sixth problem, and is still far from completion.

Over the past 30 years, significant progress in deriving hydrodynamic limits has been achieved in the realm of lattice gases with stochastic microscopic dynamics \cite{bib:Spohn2,bib:KL,bib:Liggett,bib:KLS,bib:Presutti,bib:Derrida,bib:SZ,bib:Schutz,bib:BE,bib:KRB-N,bib:Derrida07,bib:BD,bib:BD-SGJ-LL,bib:BSGJ-LL}. In particular, for stochastic lattice gases with simple conservative interactions such as exclusion processes it was shown \cite{bib:Spohn2,bib:KL} that a coarse-grained density $ \rho (x, t) $ satisfies the macroscopic conservation equation, $ \partial_t \rho = - \partial_x J$, with local current $J(x,t)$ given by Fick's law, $J = - D (\rho ) \partial_x \rho$. The diffusivity $D(\rho)$ depends on the microscopic dynamical rules and calculating, it is a very challenging problem. 

The hydrodynamic limit describes the deterministic evolution of the density field $ \rho (x, t) $ defined as a local, coarse-grained empirical average over microscopic configurations. This is analogous to the law of large numbers \cite{bib:Presutti}. The next step is to investigate fluctuations around the average, i.e., to find a property analogous to the central limit theorem for interacting lattice gases. A non-rigorous but physically well-motivated approach is to consider the density and current as stochastic fields coupled by mass conservation, $\partial_t \rho = - \partial_x J$, and to add a random contribution to the constitutive equation that relates current to density:
\begin{equation}
 J = - D (\rho ) \partial_x \rho + \sqrt{ \sigma(\rho) } \xi(x,t) \,.
\label{SPDE}
\end{equation}
Here $\xi(x,t)$ is a Gaussian white noise. The amplitude of the noise depends on the second transport coefficient $\sigma(\rho)$ known as conductivity (or mobility). Similarly to the diffusivity, the conductivity is a function of the local density. The conductivity $\sigma(\rho)$ depends on the microscopic rules. The two transport coefficients $D(\rho)$ and $\sigma(\rho)$ are difficult to calculate. If one of them is known, however, it is usually simple to determine the other due to the Einstein relation. For lattice gases close to equilibrium, the Einstein relation acquires a simple form
\begin{equation}
 \frac{2 D(\rho)}{\sigma(\rho)} = \frac{d^2 \mathcal{F}(\rho)}{d \rho^2} \,,
\label{EinsteinRelation}
\end{equation}
where $\mathcal{F}(\rho)$ is the equilibrium free energy density per site. 

Although a rigorous derivation of the stochastic partial differential equation obtained from \eqref{SPDE} is lacking, the predictions agree with available exact results established for special lattice gases, see e.g. \cite{bib:Derrida07,bib:BD}. There are also independent mathematical arguments supporting the large deviation principle implied by the stochastic Langevin equation based on \eqref{SPDE}. This is the starting point of the macroscopic fluctuation theory that describes diffusive interacting particle models in infinite domains and in finite systems connected to reservoirs at different temperatures (or chemical potentials) \cite{bib:BD-SGJ-LL,bib:BSGJ-LL}. As long as the local equilibrium is satisfied, the macroscopic fluctuation theory is applicable to far-from-equilibrium regimes. 

Thus, an understanding of macroscopic behaviors, both the deterministic (hydrodynamic) part and fluctuations around it, requires knowledge of two transport coefficients: $D(\rho)$ and $\sigma(\rho)$. The Einstein relation \eqref{EinsteinRelation} implies that it suffices to determine one coefficient. We focus on $D(\rho)$ which is especially important since it governs the hydrodynamic behavior. 

In recent years, a significant effort has been devoted to the calculation of transport coefficients for various lattice gases. When a stochastic lattice gas satisfies a special property known as the gradient condition \cite{bib:Spohn2,bib:KL}, the computations become feasible. The gradient property states that the microscopic current is the gradient of a local function (i.e., loosely speaking, Fick's law is already valid at the microscopic level). The simplest lattice gas obeying the gradient property is a collection of non-interacting random walkers---in this case, $D=1$ and $\sigma=2\rho$. The simplest interacting gradient lattice gas is the symmetric simple exclusion process \cite{bib:Spohn2,bib:Spohn} for which $D = 1$ and $\sigma = 2 \rho(1-\rho)$. Other gradient lattice gases for which the diffusivity has been computed include the Katz-Lebowitz-Spohn model with symmetric hopping \cite{bib:KLS,bib:HKPS,bib:BKL}, repulsion processes \cite{bib:Krapivsky}, a lattice gas of leap-frogging particles \cite{bib:CCGS,bib:GK}, and an exclusion process with avalanches \cite{bib:EPA}. In these models, an exact expression for the diffusivity can also be derived by a ``perturbation approach'': one writes the current at the discrete lattice level and performs a continuous limit assuming the density field to be slowly varying. 
 
Generic interacting particle processes do {\it not} satisfy the gradient condition; the computation of the diffusivity in such gases appears intractable. Nevertheless, there exists an exact variational formula for the diffusivity $D(\rho)$ derived by Varadhan and Spohn \cite{bib:Spohn} (see also \cite{bib:KLO,bib:KLO2,bib:KL}). This rather abstract formula is valid for general lattice gases regardless of the gradient condition. It expresses $D(\rho)$ as a minimum of a functional over certain classes of functions. In simple cases, the functional is quadratic and the minimization gives a set of linear equations; generally it is unclear how to determine the minimum because the function space is infinite-dimensional. 

In this work we demonstrate that the abstract variational formula for the diffusivity can be used as a tool to derive explicit (albeit approximate) formulas. We implement a systematic approximation procedure for $D(\rho)$ by minimizing over finite-dimensional subspaces of the infinite-dimensional function space. The simplest approximation gives exact results for gradient lattice gases. For general lattice gas, this iterative scheme can be carried out analytically as far as one wishes, although the complexity of calculations increases rapidly with the dimensionality of the subspace. The precision greatly improves after each step. 

The general idea of approximately solving a variational problem is widely known in science and engineering. For instance, it underlies the Ritz method. In quantum mechanics, it is known as the variational method used, e.g., in finding approximations to the ground state, and also to excited states. This venerable idea has not yet been applied to the Varadhan-Spohn formula for the diffusivity, mostly because it is little known, and it has the reputation of being a very abstract object that makes concrete calculations difficult. Furthermore, an approximate variational procedure based on the Varadhan-Spohn formula requires very long calculations even for the simplest nongradient lattice gases. Therefore one would like to choose a nongradient lattice gas which is natural and sufficiently simple to be amenable to analysis. Exclusion processes (lattice gases with at most one particle per site) appear to be good candidates. They are widely used in conceptual developments such as testing far-from-equilibrium behaviors (see \cite{bib:Spohn,bib:Derrida07,bib:BD,bib:BD-SGJ-LL,bib:BSGJ-LL}), and in various applications (see \cite{bib:CSS,bib:CKZ} and references therein). The basic example, namely the symmetric simple exclusion process, is gradient. In nongradient exclusion processes, the range of hopping is increased or there are interactions between particles occupying neighboring sites. In such situations the dimensionality of the subspace also increases---this makes computations unwieldy, and even more so if equilibrium is not given by a product measure \cite{prod_measure}.

A rather simple nongradient lattice gas with product measure is the 2-GEP, a generalized exclusion process (GEP) in which each site can host at most two particles. The 2-GEP is additionally parametrized by hopping rates depending on the occupancy levels. The 2-GEP is the first member in the family of $k$-GEP, where $k\geq 2$ is the 
maximal occupancy. The $k$-GEPs are nongradient for all $k\geq 2$ and for generic hopping rates; the notable exception is the misanthrope process \cite{bib:Cocozza-Thivent,bib:AM}. 

The GEPs have been investigated in a number of studies \cite{bib:KL,bib:KLO,bib:KLO2,bib:Seppalainen,bib:BNCPV,spider1d}. We overview their basic properties in Sec.~\ref{sec:GEPs}. In Sec.~\ref{sec:variational} we present the Varadhan-Spohn formula for the $k$-GEPs and we develop an iterative procedure allowing us to find increasingly better upper bounds for diffusivity. For the 2-GEP, each iteration of the variational procedure improves the precision by an order of magnitude. In Sec.~\ref{sec:GEPopen}, we investigate the 2-GEP on a finite interval with open boundaries connected to reservoirs. We conclude in Sec.~\ref{sec:conclusions}.

\section{Generalized exclusion processes}
\label{sec:GEPs}

We now define generalized exclusion processes (GEPs) and recall some of their basic properties at equilibrium. We start with the general $k$-GEP and then discuss the 2-GEP which is the main focus of our study. In the following we consider lattice gases in one dimension and always assume that particles undergo nearest-neighbor symmetric hopping.

\subsection{Definition of $k$-GEPs}

\begin{figure}
\begin{center}
 \includegraphics[width=80mm]{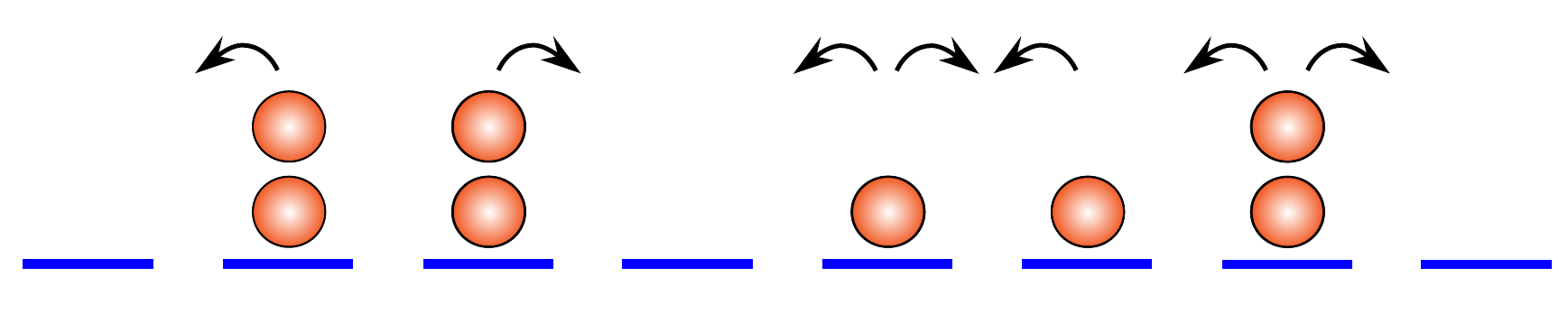}
 \caption{  
 Illustration of the GEP in one dimension, where each site is occupied by at most two particles. The arrows indicate possible transitions with rates \eqref{eq:jump}. Two particles occupying the same site jump independently,
 each with a rate $\frac{1}{2}p_{2s}$.\label{fig:GEP}}
\end{center}\end{figure}

For the $k$-GEP, each site can accommodate at most $k$ particles (see Fig.~\ref{fig:GEP}), i.e. the local occupation variables satisfy $\tau_i \in\{0,1,\ldots,k\} $. Jumps between adjacent sites are allowed only to sites with fewer than $k$ particles and the rates can depend on the occupancies of the sites:
\begin{equation}
\label{eq:jump}
\begin{split}
 & (\tau_i,\tau_{i+1}) = (r,s) \to (r-1,s+1) \quad (\text{rate}\ p_{rs}), \\
 & (\tau_i,\tau_{i+1}) = (r,s) \to (r+1,s-1) \quad (\text{rate}\ p_{sr}) . 
\end{split}
\end{equation}
The hopping is possible only from occupied sites and impossible into maximally occupied sites, so
\begin{align}
\label{eq:0rates}
 p_{ 0s } = p_{ rk } = 0 . 
\end{align}

The one-site probabilities in the equilibrium state are conveniently expressed through the fugacity $ \lambda $, 
\begin{equation}
\label{weights}
 \mathbb P [ \tau_i = r ] = W_r =\frac{ a_r \lambda^r }{Z}\,,
\end{equation}
where $Z$ is a normalization factor 
\begin{equation}
\label{Z}
 Z = \sum_{ 0 \le r \le k } a_r \lambda^r. 
\end{equation}
Without loss of generality we can set $ a_0=a_1=1 $, so that the fugacity is defined by the ratio $ W_1 / W_0 =\lambda $. 
The coefficients $a_r$ depend on hopping rates \eqref{eq:jump}. Note that the fugacity is implicitly determined by the density 
\begin{equation}
\label{rho:Z}
 \rho = \langle \tau_i \rangle = \lambda \frac{d}{d\lambda}\ln Z \,.
\end{equation}

In many GEPs, equilibrium is characterized by a product measure. For such lattice gases, the compressibility is defined by 
$ \chi = \langle \tau_i^2 \rangle - \rho^2$, and it can be alternatively written as 
\begin{align}
\label{eq:chi}
 \chi = \langle \tau_i^2 \rangle - \rho^2 = \lambda \frac{d\rho}{d\lambda} . 
\end{align}

For lattice gases with equilibrium having the product structure, the equilibrium free energy density per site also admits a simple general form (see e.g. \cite{Godreche}) 
\begin{equation}
\mathcal{F} =\rho\ln\lambda - \ln Z \,. 
\label{FreeE}
\end{equation}
Differentiating $\mathcal{F}$ with respect to $\rho$ and using \eqref{rho:Z} we obtain $\frac{d \mathcal{F}}{d \rho}=\ln\lambda$. Differentiating again gives $\frac{d^2 \mathcal{F}}{d \rho^2}=\frac{1}{\chi}$, so that the Einstein relation \eqref{EinsteinRelation} can be rewritten as
\begin{equation}
\label{ER}
\sigma = 2\chi D. 
\end{equation}

\subsection{2-GEP}

For the 2-GEP, there are four independent generally non-vanishing hopping rates: $p_{10} , p_{11} , p_{20} , p_{21}$. The remaining five vanish: $p_{ 00 } = p_{ 01 } = p_{ 02 } = p_{ 12 } = p_{ 22 } = 0$.

Some special sets of rates have been considered in the literature, particularly
\begin{itemize}
 \item \textit{Particle-uniform rates}: Each particle hops with the same (unit) rate as long as the maximal occupancy constraint is obeyed (see e.g. \cite{bib:AKM}). Therefore the non-vanishing rates are 
\begin{align} 
\label{eq:rates}
 p_{20} = p_{21} = 2, \quad p_{10} = p_{11} = 1. 
\end{align}
\item \textit{Site-uniform rates}: The occupancy of each site is updated with rate 1 as 
as long as the maximal occupancy constraint is obeyed (see e.g. \cite{bib:KL,bib:Seppalainen,spider1d})
\begin{align} 
\label{eq:another}
 p_{20} = p_{21} = p_{10} = p_{11} = 1. 
\end{align}
\item The \textit{misanthrope process} \cite{bib:Cocozza-Thivent}:
\begin{align} 
\label{eq:misanthrope}
 p_{20} = p_{21} + p_{10} . 
\end{align}
\end{itemize}
Whenever possible we shall consider the general 2-GEP with parameters $p_{10} , p_{11} , p_{20} , p_{21}$ being arbitrary non-negative numbers, but some results valid specifically for the rates \eqref{eq:rates}, \eqref{eq:another} or \eqref{eq:misanthrope} will be emphasized. In particular, simulations have been performed for the 2-GEP with particle-uniform rates \eqref{eq:rates}. 

We have three one-site probabilities in the equilibrium
\begin{equation}
\label{eq:single-weights}
 W_0 = \frac{1}{Z} \,, ~
 W_1= \frac{\lambda}{Z} \,, ~
 W_2 = a \frac{ \lambda^2}{ Z}\, 
\end{equation}
with $ a_2 =a $ for simplicity. The normalization factor \eqref{Z} turns into a quadratic polynomial 
\begin{equation}
\label{eq:Z=}
Z= 1 + \lambda + a \lambda^2 
\end{equation}
 for the 2-GEP. The product measure holds for the 2-GEP, viz. the probability of each configuration takes the product form \cite{bib:KL} at equilibrium. In particular, 
\begin{equation}
\label{prod}
\mathbb P [ \tau_i = r,\, \tau_{i+1} = s ] = W_r W_s\,.
\end{equation}
The validity of the product measure can be seen by checking the detailed balance condition corresponding to the processes $(r,s)\leftrightarrow (r-1,s+1)$: 
\begin{equation}
\label{DB}
p_{rs} W_r W_s= p_{r+1,s-1} W_{s-1} W_{r+1}\,.
\end{equation}
For the 2-GEP, these equations are identities for most $(r,s)$, the only exception is 
\begin{equation}
\label{eq:aX1X1-X0X2=0} 
 p_{20} W_2 W_0= p_{11} W_1^2\, , 
\end{equation}
which fixes the coefficient $a $ in \eqref{eq:single-weights} to be 
\begin{equation}
\label{eq:a=}
 a=\frac{ p_{11} }{ p_{20} }\, . 
\end{equation}
The single-site probabilities $W_r$, and generally the probabilities of arbitrary configurations which are the products of single-site probabilities, could have been functions of $p_{10} , p_{11} , p_{20} , p_{21}$.
However, they actually depend only on the ratio \eqref{eq:a=}. 

For the 2-GEP, the density and the compressibility can be written as 
\begin{align}
\label{eq:rho=W1+2W2}
 \rho =& W_1 + 2 W_2 ,\\
\label{eq:chi-weights} 
 \chi =& W_1 + 4 W_2 - \left(W_1 + 2 W_2\right)^2 . 
\end{align}
Combining \eqref{eq:single-weights} and \eqref{eq:rho=W1+2W2} we determine an explicit formula for $ \lambda $ in terms of $ \rho $
\begin{align}\label{eq:lambda=}
\lambda = \frac{ 2 \rho }{ 1- \rho + \sqrt{1-(1-4a) \rho(2-\rho) } }
\end{align}
and then we express the compressibility via density:
\begin{align}
\chi = \frac{ \rho (2-\rho) \sqrt{1-(1-4a) \rho(2-\rho) } }{ 1+\sqrt{1-(1-4a) \rho(2-\rho) } }\, . 
\end{align}

We always assume that $0<a<\infty$. Peculiar behaviors may occur in the extreme cases of $a=0$ and $a=\infty$. These subtleties are outlined in Appendix~\ref{app:pecular}. Possible qualitative differences between the 2-GEP and $k$-GEPs with $k\geq 3$ are outlined in Appendix~\ref{app:3-GEP}.

\section{Bulk diffusivity: A variational calculation}
\label{sec:variational}

\subsection{Varadhan-Spohn formula for the diffusivity}
\label{sec:Varadhan}

We first review the Varadhan-Spohn variational formula giving the diffusion coefficient $D(\rho)$ of a stochastic lattice gas. Generally, $D(\rho)$ can be expressed, via the Green-Kubo formula, as an integral of a current-current correlation function \cite{bib:KLS,bib:Spohn2}. The Green-Kubo expression can be rewritten as the solution of a variational problem (as shown by Spohn \cite{bib:Spohn2} who attributes this variational approach to Varadhan). More precisely \cite{bib:Spohn2,bib:KLO2}:
\begin{align}
\label{eq:variational}
 D = \frac{1}{2\chi} \inf_{ f }\, \langle Q(f) \rangle\,. 
\end{align}
Here $ \chi $ is the compressibility and $ Q(f) $ is a quadratic functional of the space of functions $f$ which depend only on finite points of $ \tau $ (``cylinder'' functions). The expectation value $ \langle \cdot \rangle $ is taken with respect to the equilibrium measure on the configuration space, see Sec.~\ref{sec:GEPs}.

The precise form of the functional $ Q(f) $ depends on the microscopic dynamical rules of the process. To write down $ Q(f) $ for the $k$-GEP, it is convenient to use some auxiliary notations. For any configuration 
$\tau = ( \ldots,\tau_{-1},\tau_0,\tau_1, \tau_2, \ldots ) $ we denote by $\tau^{0 \rightarrow 1}$ the configuration obtained from $ \tau $ by making a single particle jump from site 0 to site 1 if such move move is permitted, otherwise $\tau^{0 \rightarrow 1}$ is identical to $ \tau $:
\begin{align}
\label{eq:hop01} \tau^{0 \rightarrow 1} = 
\begin{cases} 
( \ldots,\tau_0-1,\tau_{1}+1,\ldots ) &\tau_0 \ge 1, ~ \tau_{1} < k \\ 
\tau &\mbox{otherwise }. 
\end{cases}
\end{align}
When the particle hop is possible, two occupation numbers in $\tau^{0 \rightarrow 1}$ differ from the corresponding occupation numbers in $\tau$, and only these numbers are explicitly shown in  Eq.~\eqref{eq:hop01} . 
Similarly the configuration $\tau^{0 \leftarrow 1}$ is obtained from $ \tau $ by making a particle hop from site 1 to site 0 if possible:
\begin{align}
\label{hop10}
\tau^{0 \leftarrow 1} = (\ldots,\tau_0+1,\tau_{1}-1,\ldots) , 
\end{align}
when $\tau_0 < k$ and $\tau_1\ge 1$; otherwise $\tau^{0 \leftarrow 1} = \tau$. 
For any cylinder function $f$ and $ j\in \mathbb Z$,
we define $ f^j $ by 
\begin{align}
 f^{j}(\tau) = f ( T_ j \tau), 
\end{align}
where the translation operator $T_j$ shifts the configuration $\tau$ forward by $j$ sites: $(T_j \tau)_i = \tau_{i-j}.$
In other words, we shift the frame of $ f $ backward by $ j $ sites. For instance, if $ f $ depends only on the
$ (\tau_1,\tau_2\cdots,\tau_n) $, that is, 
\begin{align}
 f (\tau) = f ( \tau_1,\tau_2\cdots,\tau_n ), 
\end{align}
the shifted function $f^j$ acts on the configuration space as 
\begin{align}
\label{eq:fjtau=f(tau...tau)}
 f^{j}(\tau) = f ( \tau_{1-j},\tau_{2-j}\cdots,\tau_{n-j} ) . 
\end{align}
Note that $ f^0 \equiv f $.

We can now define $Q(f)$ by its action on the space of cylinder functions $f$:
\begin{align}
Q(f)(\tau) = p_{\tau_0\tau_{1}}
 \bigg[1 - \sum_{j \in \mathbb Z } \left(f^{j}(\tau^{0 \rightarrow 1}) - f^{j}(\tau)
 \right) \bigg]^2
 \nonumber \\ 
 + \, p_{\tau_1\tau_{0}}
 \bigg[ -1- \sum_{j \in \mathbb Z } \left(f^j(\tau^{0 \leftarrow 1}) - f^j(\tau) \right) \bigg]^2 . 
\label{eq:originalQf}
\end{align}

The sum with respect to $j$ in \eqref{eq:originalQf} contains a finite number of terms when the function $f$ has a finite range. To appreciate this assertion consider a function of type \eqref{eq:fjtau=f(tau...tau)}. For $j < 0 $ and $j > n $, the shifted function $f^j$ is not sensitive to the values of $\tau_0$ and $\tau_1$, i.e., in Eq. \eqref{eq:fjtau=f(tau...tau)} $\tau_0$ and $\tau_1$ do not appear. Therefore $f^j(\tau^{0 \rightarrow 1}) - f^j(\tau)=0$ for such $j$. Thanks to this property, one can replace $ \sum_{j \in \mathbb Z } $ in \eqref{eq:originalQf} by $ \sum_{ 0\le j \le n } $. Only a 
finite number of variables, viz. $ (\tau_{-n+1},\tau_{-n},\cdots, \tau_{n-1}, \tau_n) $,
appear in \eqref{eq:originalQf}.

\subsection{Iterative scheme}

The Varadhan-Spohn formula \eqref{eq:variational} is a powerful theoretical tool, but it is unclear how to apply it even to simplest lattice gases such as the 2-GEP. One can try to obtain an upper bound for the diffusivity by restricting the space of functions. Let us consider the subspace $ F_n $ of functions depending on the configuration of $ n $ adjacent sites $( \tau_1, \tau_2, \dots, \tau_n )$ and use a restricted version of the Varadhan-Spohn formula 
\begin{align}
 D_n = \frac{1}{2\chi} \min_{ f \in F_n } \langle Q(f) \rangle \,. 
\end{align}
Each subspace $F_n$ includes the previous ones, $F_{n-1}\subset F_n$, and therefore 
inequalities 
\begin{align}
 D_0 \ge D_1 \ge D_2 \ge \cdots . 
\end{align}
The series $ D_n $ converges to the actual diffusivity: 
\begin{align}
 \lim_{n\to \infty} D_n = D .
\end{align}

The subspace $F_0$ consists of constant functions. Therefore all sums in \eqref{eq:originalQf} vanish and the diffusivity in the zeroth order approximation reads
\begin{align}
\label{eq:D0=}
 D_0 = \frac{ \big\langle p_{\tau_0\tau_{1} } \big\rangle }{ \chi } . 
\end{align}
This equation is valid for $k$-GEPs with any $k$; the same hold for all previous results of this section. In the rest of this work, we consider the simplest 2-GEP models if not stated otherwise.

For the 2-GEP, the zeroth order approximation \eqref{eq:D0=} for the diffusion coefficient becomes 
\begin{equation}
\label{eq:D0:2GEP}
 D_0 = \frac{ p_{10} + 2 p_{11} \lambda + a p_{21} \lambda^2 }{ 1 + 4a\lambda + a\lambda^2 }\,. 
\end{equation}
It is also possible to obtain an explicit form of $D_0$ as a function of $\rho$ by substituting \eqref{eq:lambda=} into \eqref{eq:D0:2GEP}. We present $D_0$ in two cases. Specializing \eqref{eq:D0:2GEP} to the particle-uniform rates \eqref{eq:rates} we obtain
\begin{equation}
\label{eq:ourD0}
 D_0 = \frac{ 2 ( 1+ \lambda)^2 }{ 2 + 4 \lambda + \lambda^2 }
 = \frac{1 + \rho +\sqrt{1+2\rho-\rho^2}}{2\sqrt{1+2\rho-\rho^2}} 
\end{equation}
while in the site-uniform case \eqref{eq:another} \cite{bib:KLO} 
\begin{align}
\label{eq:otherD0}
 D_0 = \frac{ ( 1+ \lambda)^2 }{ 1 + 4 \lambda + \lambda^2 }
 = \frac{ 2 +\sqrt{1+6\rho-3\rho^2}}{3\sqrt{1+6\rho-3\rho^2}} . 
\end{align}

There is a symmetry between particles and vacancies for the 2-GEP with site-uniform rates. This symmetry implies the mirror symmetry of the diffusivity, that is, the invariance with respect to the $\rho\leftrightarrow 2-\rho$ transformation:
\begin{equation}
\label{mirror}
D(\rho)=D(2-\rho)\,.
\end{equation}
In terms of the fugacity, the mirror symmetry reads
\begin{equation}
\label{mirror:fug}
D(\lambda)=D(1/\lambda)\,.
\end{equation}
Our iterative scheme gives approximations of the diffusivity agreeing with mirror symmetry. This is evident in the zeroth order, Eq.~\eqref{eq:otherD0}, and holds in all higher orders.

\subsection{The first-order approximation}
\label{subsec:1}

\begin{figure}
 \begin{center}
 \includegraphics[width=85mm]{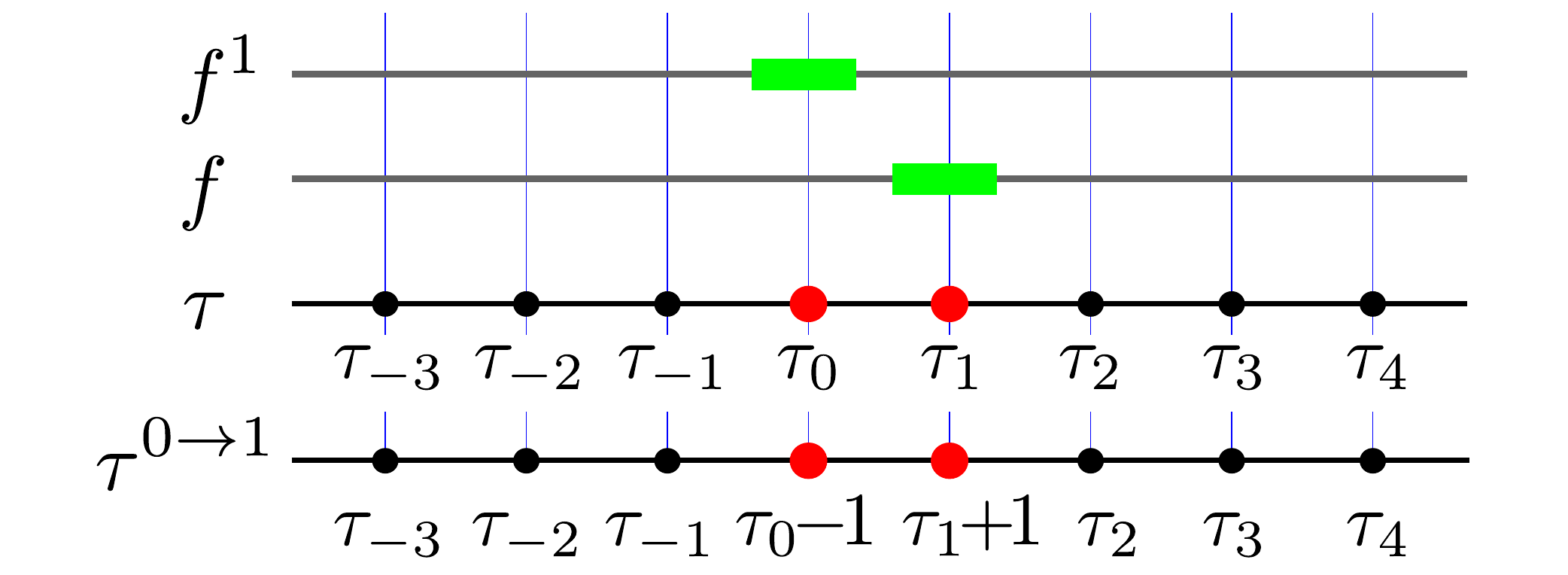}
 \caption{  Illustration of the procedure of computing $ Q(f) $ for $ n=1 $ when the local function $f(\tau)=f(\tau_1)$
 depends only on a single site and hence only $f^0\equiv f$ and $f^1$ contribute to \eqref{eq:originalQf}. \label{fig:n1}} 
\end{center}
\end{figure} 

The case of $n=1$ is more instructive, although we shall see that the final result is the same as for $n=0$. When $n=1$, the functions $f$ depend only on $\tau_1$, that is, $f(\tau)=f(\tau_1) :=f_{\tau_1}$. There are three independent values, $f_0,f_1$ and $f_2$, so $F_1$ is a three-dimensional space. 

Only $ j =0 $ and $j=1$, see Fig.~\ref{fig:n1}, contribute to the sum $\sum_j$ in Eq.~\eqref{eq:originalQf}. Writing $ \tau_{0} = r $ and $\tau_1 = s $, we have 
\begin{align}
\label{Qf:1}
 Q(f)(\tau) = \sum_{\epsilon = \pm 1 } p_{rs} ( \epsilon - f_{r-1} + f_r - f_{s+1} + f_s)^2 , 
\end{align}
where the summation over $\epsilon = \pm 1$ combines the two terms on the right-hand side of Eq.~\eqref{eq:originalQf} into a single sum. Averaging \eqref{Qf:1} with respect to the equilibrium measure we obtain
\begin{equation}
 \begin{split}
 \langle Q(f) \rangle 
 =&\sum_{ r=1,2 \atop s=0,1 } \sum_{\epsilon = \pm 1 } p_{rs} W_r W_s\\ 
 & ( \epsilon - f_{r-1} + f_r - f_{s+1} + f_s)^2 . 
 \end{split}
\end{equation}
By a straightforward calculation, one finds 
\begin{align}
\langle Q(f) \rangle = 2 \big\langle p_{rs} \big\rangle + 
 ( f_0 - 2 f_1 + f_2 )^2 \frac{ 4 \lambda^2 p_{11} }{ Z^2 } . 
\end{align}
This is minimized when $ f_0 - 2 f_1 + f_2 = 0 $ and gives $D_1=D_0$. Thus the first iteration does not lead to any improvement. 

 \subsection{The second-order approximation}
 \label{subsec:2}

\begin{figure}
 \begin{center}
 \includegraphics[width=85mm]{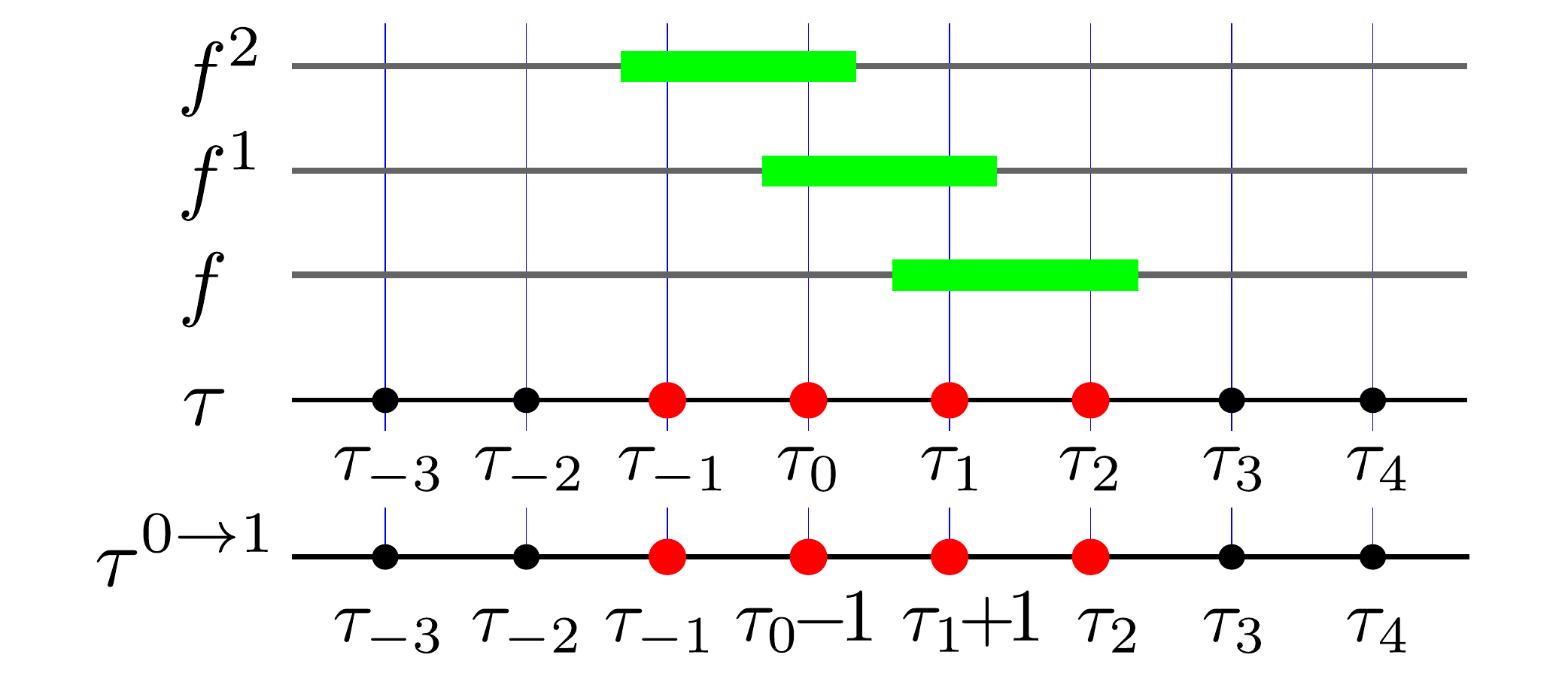}
 \caption{  Illustration of the procedure for $ n=2 $, when the local function $f$ 
 depends on two adjacent sites and only $f,~f^1$ and $f^2$ contribute to \eqref{eq:originalQf}. 
 \label{fig:n2}} 
\end{center}
\end{figure} 

Now $ f(\tau) = f(\tau_0,\tau_1) =: f_{ \tau_0,\tau_1 } $ can take nine different values. 
In the functional $Q(f)$, the sites $ j $ contributing to the sum $\sum_j$ in Eq.~\eqref{eq:originalQf} are $j\in \{0,1,2 \} $, 
see Fig.~\ref{fig:n2}. We write $ \tau_{-1} = q , \,\tau_0 = r , \,\tau_1 = s , \, \tau_2 = u $. The average of $Q(f)$ becomes 
\begin{align}
 & \langle Q(f) \rangle = 
 \sum_{q,r,s,u\in \{ 0,1,2 \} } 
\sum_{ \epsilon =\pm 1 }
 p_{rs} W_q W_r W_s W_u \\ 
 & (\epsilon - f_{q,r-1} + f_{q,r} - f_{ r-1,s+1 } + f_{r,s} - f_{s+1,u} + f_{s,u} )^2. \nonumber
\end{align}

To determine the minimum, we must solve $ \frac{ \partial \langle Q(f) \rangle }{ \partial f_{ \xi \eta } } =0 $ ($ \xi , \eta \in \{ 0,1,2 \} $). There are five independent equations characterizing the solution space.
Four of them are homogeneous
\begin{equation}
\begin{split}
\label{ff:many}
 f_{20} - 2 f_{11} & = 0 , \\
 f_{10} - f_{00} - f_{11} & = 0 , \\
 f_{22} + f_{00} - 2f_{11} & = 0 , \\
 f_{21} + f_{12} + f_{10} - f_{20} - 2f_{11} & = 0 . 
\end{split}
\end{equation}
An additional inhomogeneous equation is 
\begin{equation}
\label{ff}
 f_{12} + f_{01} - f_{02} - f_{11} = \frac{p_{21} + p_{10} - p_{20}}{V}
\end{equation}
with 
\begin{equation*}
V = p_{21} + p_{10} + p_{20} + \frac{p_{21} +2 p_{11} \lambda + p_{10} a \lambda^2}{ Z } \,.
\end{equation*}
Using \eqref{ff:many}--\eqref{ff} we find the minimum of 
$\langle Q(f) \rangle$, from which 
\begin{equation}
\label{D2:2GEP}
 D_2 = D_0 - \frac{ 2 ( p_{21} + p_{10} - p_{20} )^2 \lambda^2 }{ ( a^{-1} + 4 \lambda + \lambda^2 ) V Z } . 
\end{equation}
The second order approximation gives a better prediction for the diffusion coefficient: $ D_2< D_1= D_0 $. The only exception is the misanthrope process \eqref{eq:misanthrope} which is gradient and therefore the zeroth order approximation is already an exact answer. 

\begin{figure}
 \begin{center}
 \includegraphics[width=60mm]{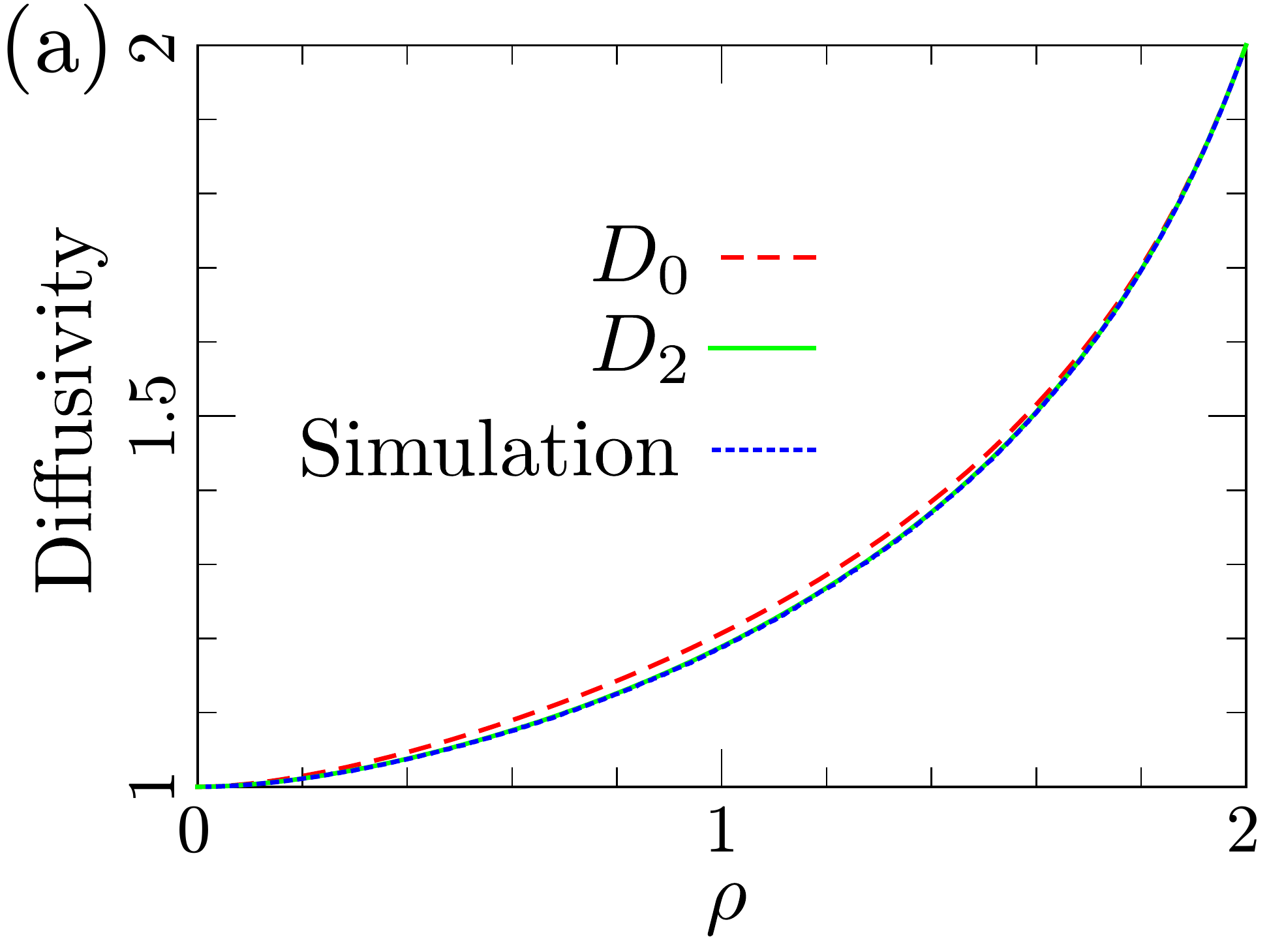}
 \includegraphics[width=60mm]{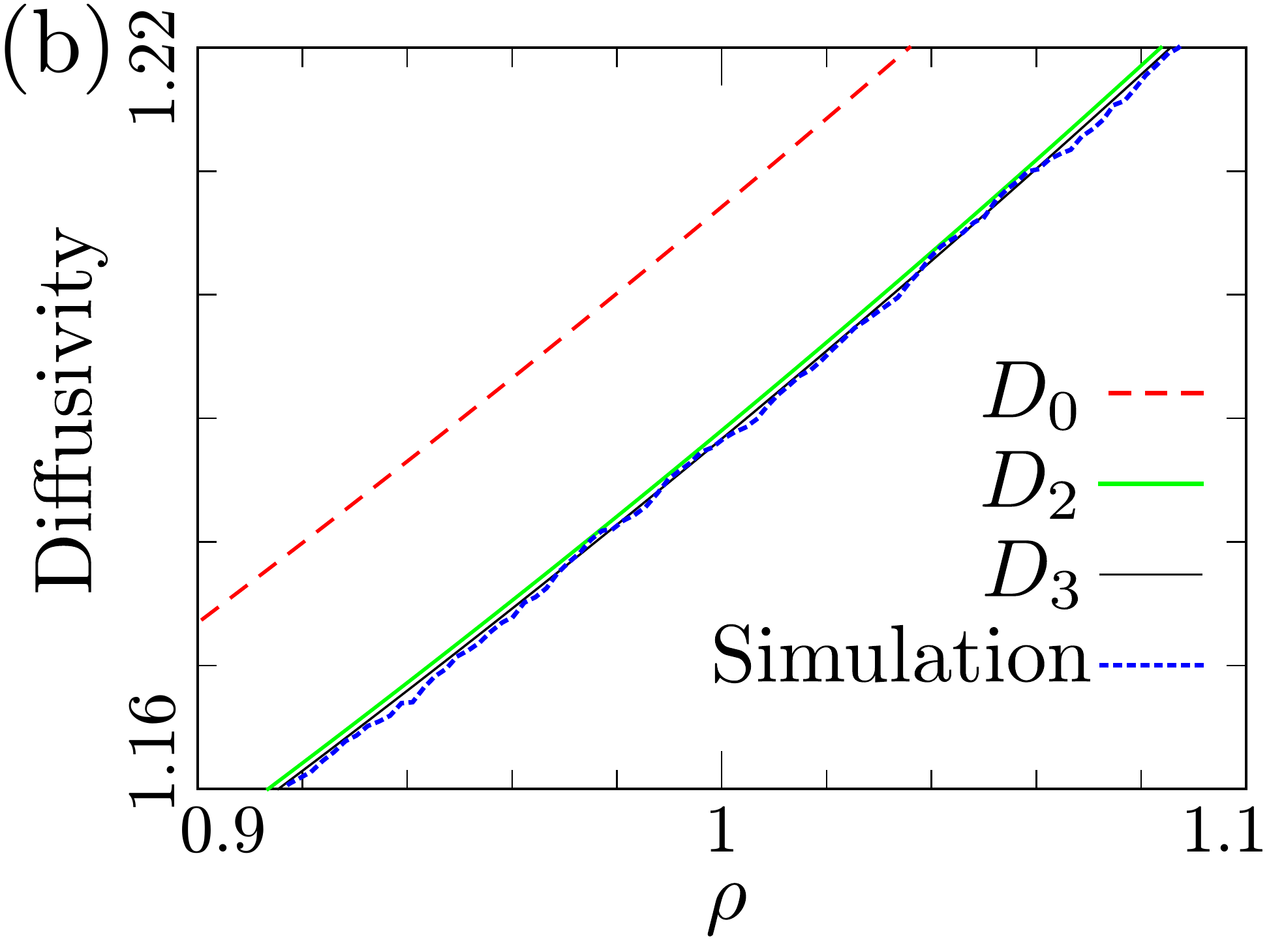}
 \caption{(a) Zeroth- and second-order approximations of the diffusivity over the entire density range.  (b) Zeroth-, second, and third-order approximations in the density range $0.9\leq \rho\leq 1.1$.  Numerical results (dotted lines) were obtained by simulating an open system of length $ L=1024 $. 
 \label{fig:D0D2D3sim}} 
\end{center}
\end{figure}

Specializing the general expression \eqref{D2:2GEP} to the particle-uniform rates \eqref{eq:rates} we obtain 
\begin{align}
\label{eq:D2=}
 D_2 = \frac{ 2(7+21\lambda+23\lambda^2+13\lambda^3+3\lambda^4) }
 {(2+4\lambda+\lambda^2)(7+7\lambda+3\lambda^2)} . 
\end{align}
Figure \ref{fig:D0D2D3sim} (a) shows $D_0(\rho) , D_2(\rho) $ and simulation results obtained for the system with open boundaries (Sec.~\ref{sec:GEPopen}). We see that simulation data almost perfectly match analytical results for $ D_2 $. 

For the site-uniform rates \eqref{eq:another} we get
\begin{align}
 D_2 = \frac{ 4 + 13\lambda + 16 \lambda^2 + 13 \lambda^3 + 4\lambda^4 }{
 (1+4\lambda+\lambda^2) (4+5\lambda+4\lambda^2) }\, . 
\end{align}
The second-order approximation obeys the mirror symmetry \eqref{mirror:fug} as expected.

 \subsection{The third and higher order approximations}
 \label{sub:3}

\begin{figure}
\begin{center}
 \includegraphics[width=85mm]{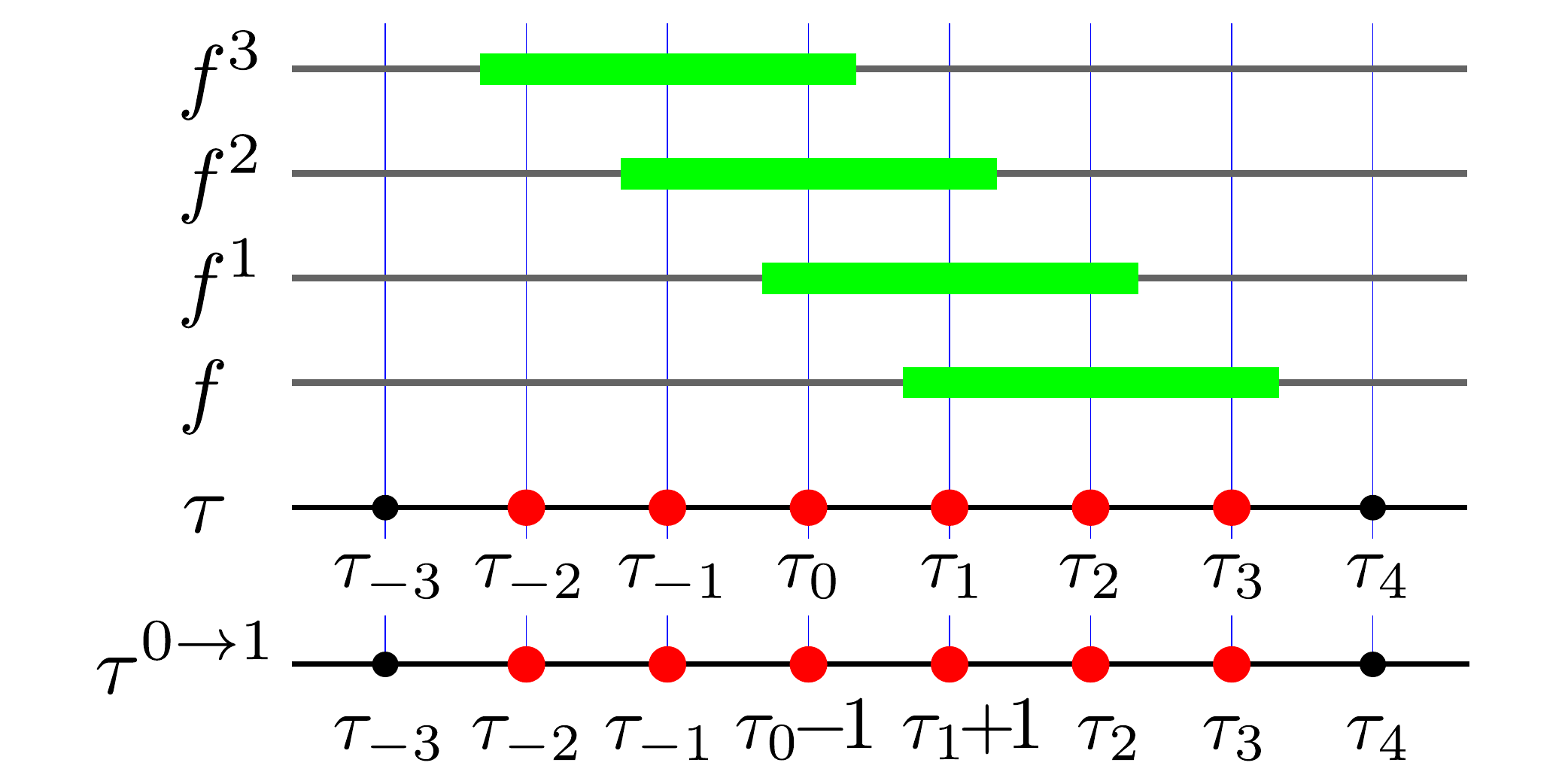}
 \caption{  Illustration of the procedure for $ n=3 $, when the local function $f$ 
 depends on three adjacent sites and only $f,~f^1, ~f^2$ and $f^3$ contribute to \eqref{eq:originalQf}. \label{fig:n3}} 
\end{center}
\end{figure}

In the third order, $ j \in \{0,1,2,3 \} $ contribute to the sum $\sum_j$ in Eq.~\eqref{eq:originalQf}, see Fig.~\ref{fig:n3}. We denote by $ \tau_{-2} = \ell, \,\tau_{-1} = q , \,\tau_0 = r , \,\tau_1 = s , \, \tau_2 = u , \, \tau_3 = v $ the relevant occupation numbers and recast $\langle Q(f) \rangle$ into 
\begin{align}
\nonumber
 & \langle Q(f) \rangle = 
 \sum_{\ell,q,r,s,u,v\in \{ 0,1,2 \} } 
\sum_{ \epsilon =\pm 1 }
 p_{rs} W_\ell W_q W_r W_s W_u W_v \\ 
\begin{split}
 & (\epsilon - f_{\ell,q,r-1 } + f_{\ell, q, r } - f_{q,r-1,s+1} + f_{q,r,s} \\
 & - f_{ r-1,s+1,u } + f_{r,s,u} - f_{s+1,u,v} + f_{s,u,v} )^2.
\end{split}
\end{align}
We must solve $ \frac{ \partial \langle Q(f) \rangle }{ \partial f_{\xi\eta\zeta} } = 0 $ ($ \xi , \eta,\zeta \in \{ 0,1,2 \} $). There are 17 independent equations which are listed in Appendix~\ref{sec:n=3}. 

Lengthy but straightforward calculations give 
\begin{equation}
\label{D3:AB}
 D_3 = \frac{ a^{-1} A(\lambda) }{ ( a^{-1} + 4 \lambda + \lambda^2 ) B(\lambda) }\,. 
\end{equation}
where $A(\lambda)$ and $B(\lambda)$ are polynomials in $ \lambda $ of the 8th and 6th degree, respectively. We do not display long explicit formulas for $A$ and $B$ for the general rates $ p_{rs} $ and limit ourselves with two examples. 

For the particle-uniform rates \eqref{eq:rates}, we have $a=\frac{1}{2}$ and 
\begin{eqnarray*}
A(\lambda) &=& 2870+16194 \lambda +40422 \lambda^2+59250 \lambda^3 + 56100 \lambda^4\\
 &+& 35274 \lambda^5+14410 \lambda^6 +3499 \lambda^7 +385 \lambda^8\,,\\
B(\lambda) & = & 2870+10454 \lambda +17064 \lambda ^2+15772 \lambda ^3 + 8707 \lambda ^4 \\
 & + & 2729 \lambda ^5+385 \lambda ^6 .
\end{eqnarray*}
For the site-uniform rates \eqref{eq:another}, we have $a=1$ and 
\begin{eqnarray*}
A(\lambda) &=& 225+ 1428\lambda + 4140\lambda^2+7396\lambda^3 + 8886\lambda^4\\ 
 & + & 7396\lambda ^5+ 4140\lambda^6 + 1428\lambda^7+225 \lambda^8, \\
 B(\lambda) &=& 225 + 978\lambda +2079 \lambda ^2+2620 \lambda ^3 +2079 \lambda ^4\\ 
 & + & 978 \lambda ^5+225 \lambda^6 . 
\end{eqnarray*}

\begin{figure}
\begin{center}
 \includegraphics[width=60mm]{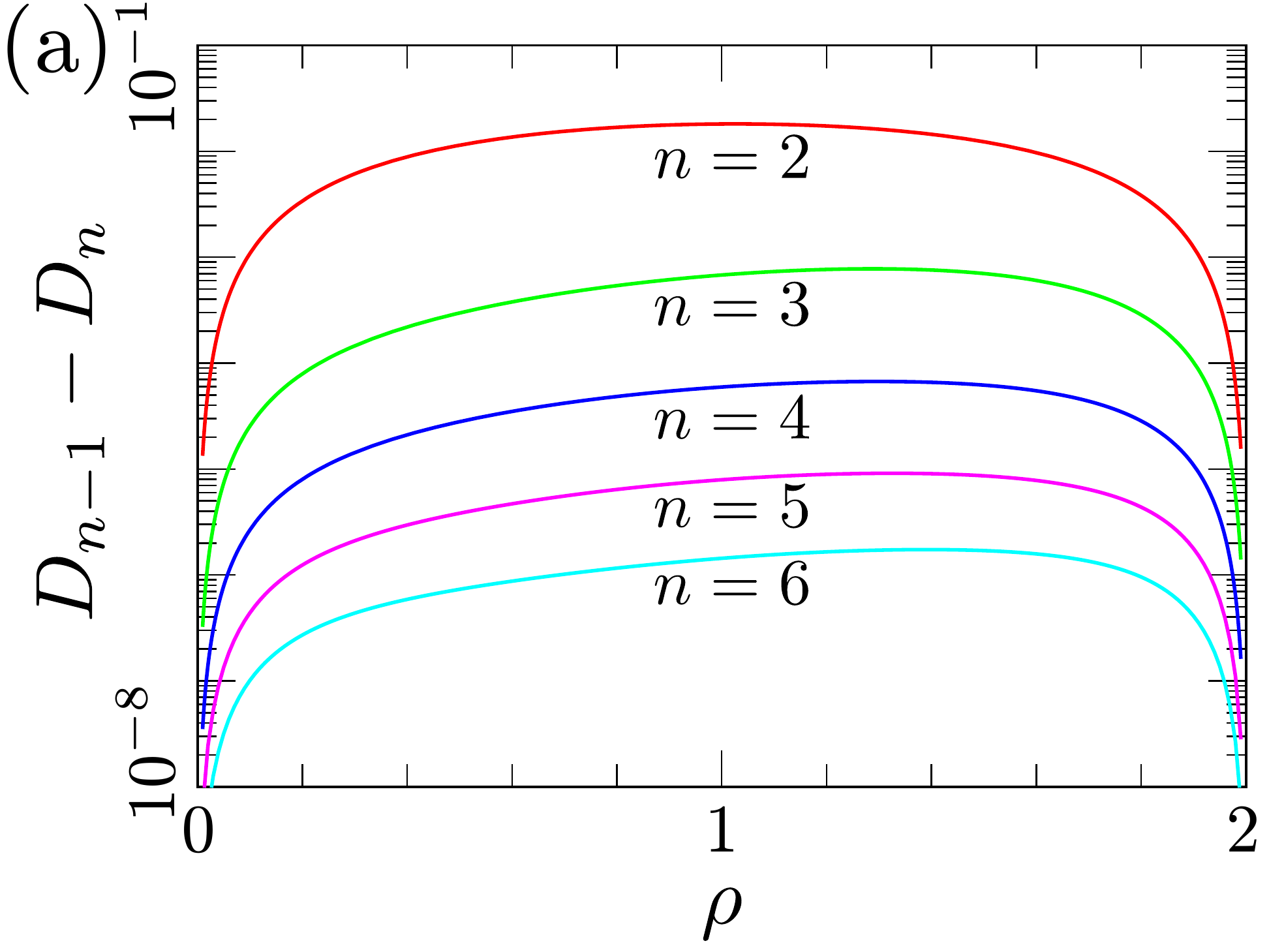}
 \includegraphics[width=60mm]{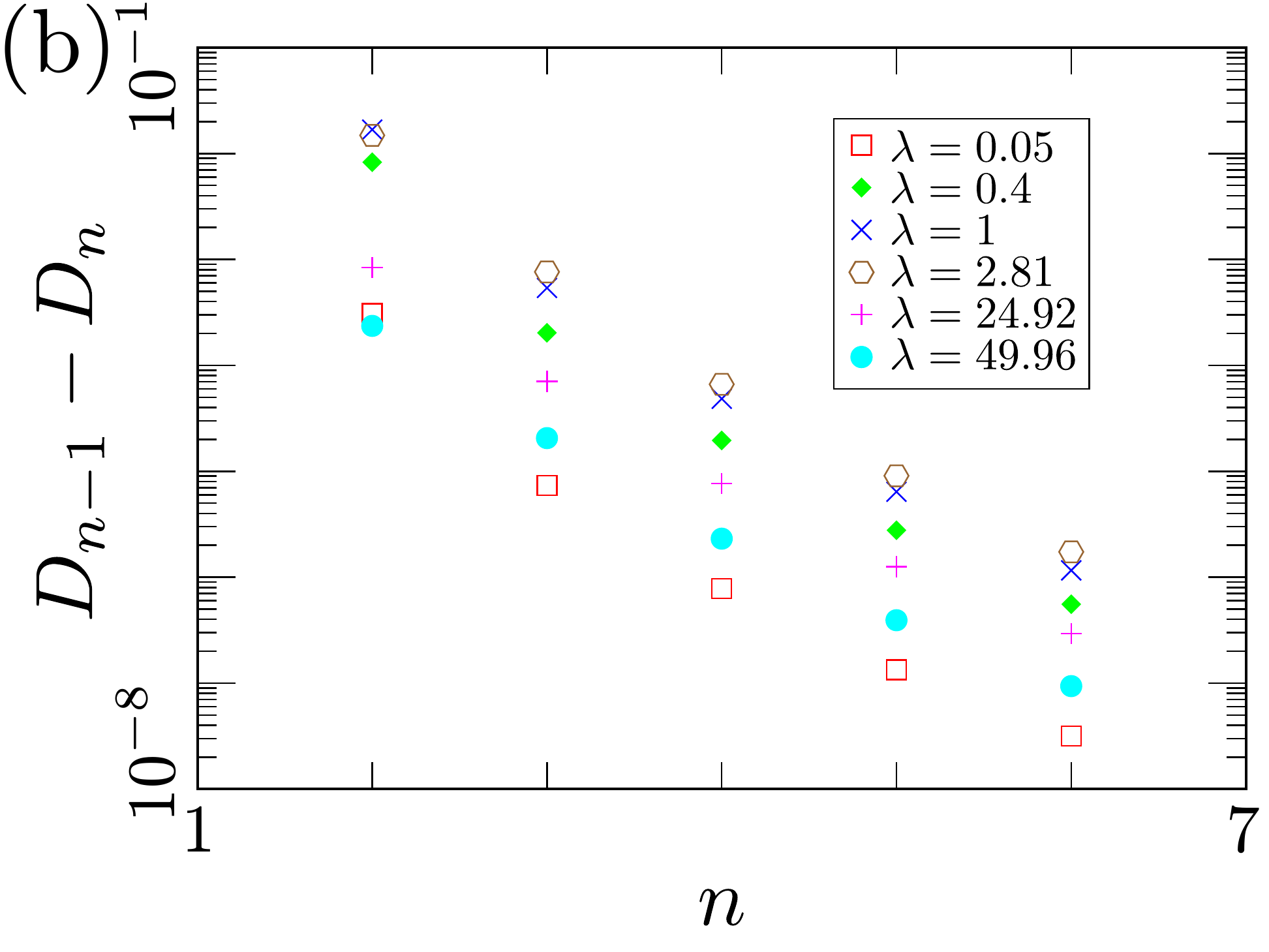}
 \caption{  Difference between $n$th and $(n-1)$st order approximations vs $\rho$ (a) and $n$ (b). The plots are for the 2-GEP with particle-uniform rates \eqref{eq:rates}. The quality of approximation is the worst around half-filling; in the examples of (b), the quality is worst when $\lambda=2.81$ corresponding to $\rho=1.38$. \label{fig:D-D}}
\end{center}
\end{figure} 

In Fig.~\ref{fig:D0D2D3sim} (b), we plot the curves $ D_0 $, $D_2 $ and $ D_3 $ as well as the simulation result around $ \rho =1 $. We find that the difference between $ D_2$ and $ D_3 $ is almost the same magnitude of the statistical error of the simulation.

It is possible to determine analytical expressions for $D_n$ for any finite $ n $, but calculations become cumbersome as $n$ increases. For $n=4$, for instance, even in a simple case of particle-uniform rates the analytical expression for $D_n$ has the form \eqref{D3:AB}, with $A(\lambda)$ and $B(\lambda)$ being polynomials with integer coefficients of degree 30 and 28, respectively (the coefficients are huge, some are of the order of $10^{24}$). Figure \ref{fig:D-D} shows $ D_{n-1} - D_n$ for $n \leq 6$ for the case of the particle-uniform rates \eqref{eq:rates}. The difference $ D_5-D_4 $, for instance, is less than $ 10^{-5} $ in the entire density region $ 0 < \rho < 2 $. We expect that $ D_{n-1} - D_n$ decreases faster than algebraically. 

In a previous work \cite{bib:AKM}, we calculated the diffusion coefficient for $k$-GEPs using an approach that actually gave the zeroth-order approximation, Eq.~\eqref{eq:D0=}. This prediction was in good agreement with Monte Carlo simulations, but further investigations \cite{bib:BNCPV2,bib:AKM2} of the 2-GEP revealed a discrepancy between the prediction and the actual value. The above approach shows how to improve this approximation in a systematic manner.

\section{GEPs with open boundaries}
\label{sec:GEPopen}

On a ring, the 2-GEP and some classes of the $k$-GEPs (including particle-uniform and site-uniform cases) satisfy detailed balance, and equilibrium steady states are described by a product measure \cite{bib:Cocozza-Thivent}. On an open chain connected to reservoirs at different densities, however, these processes are non-equilibrium   and their steady states are generally unknown. (The simple exclusion process $k=1$ is an exceptional case where the steady state is known, see e.g. \cite{bib:BE,bib:Derrida07}.) In the large system size limit, the deviation from equilibrium is small and one can use Fick's law to describe the hydrodynamic behavior \cite{bib:KL}.  In this section we study numerically more subtle properties such as finite-size corrections to the product measure.

\begin{figure}\begin{center}
 \includegraphics[width=80mm]{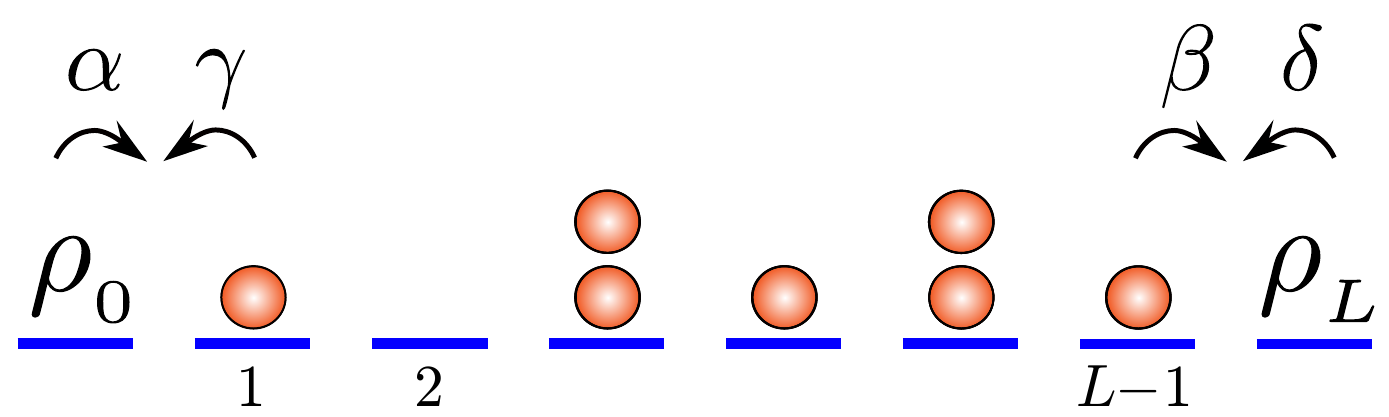}
 \caption{  
Illustration of the 2-GEP with open boundaries. 
The ``sites'' 0 and $L$ correspond to reservoirs with (in general) different densities. 
These boundary conditions are realized by transitions of the states at sites 1 and $ L-1 $, 
which are indicated by arrows. 
 \label{fig:open}}
\end{center}\end{figure}

Specifically, we consider the 2-GEP on a finite chain with $ L-1 $ bulk sites connected to reservoirs with densities $ \rho_0 $ and $ \rho_L $, see Fig.~\ref{fig:open}. The couplings to the reservoirs are described by injection and extraction rates $ \alpha , \beta, \gamma , \delta $, see Fig.~\ref{fig:open}. These rates are determined by the densities $ \rho_0 , \rho_L $ and the bulk hopping rates, see Appendix~\ref{sec:app-coupling} for details.
We use simulations and scaling considerations to shed light on non-equilibrium steady states. We perform simulations for the 2-GEP with particle-uniform rates. In our simulations, we consider extreme boundary densities, $ \rho_0 = 2$ and $ \rho_L = 0$. Further, we perform a time average over $ 10^7 \le t \le 10^9 $ and an ensemble average over 10 independent runs. 

\subsection{Density profiles in the open system}
\label{sec:density} 
 
\begin{figure}
\begin{center}
 \includegraphics[width=60mm]{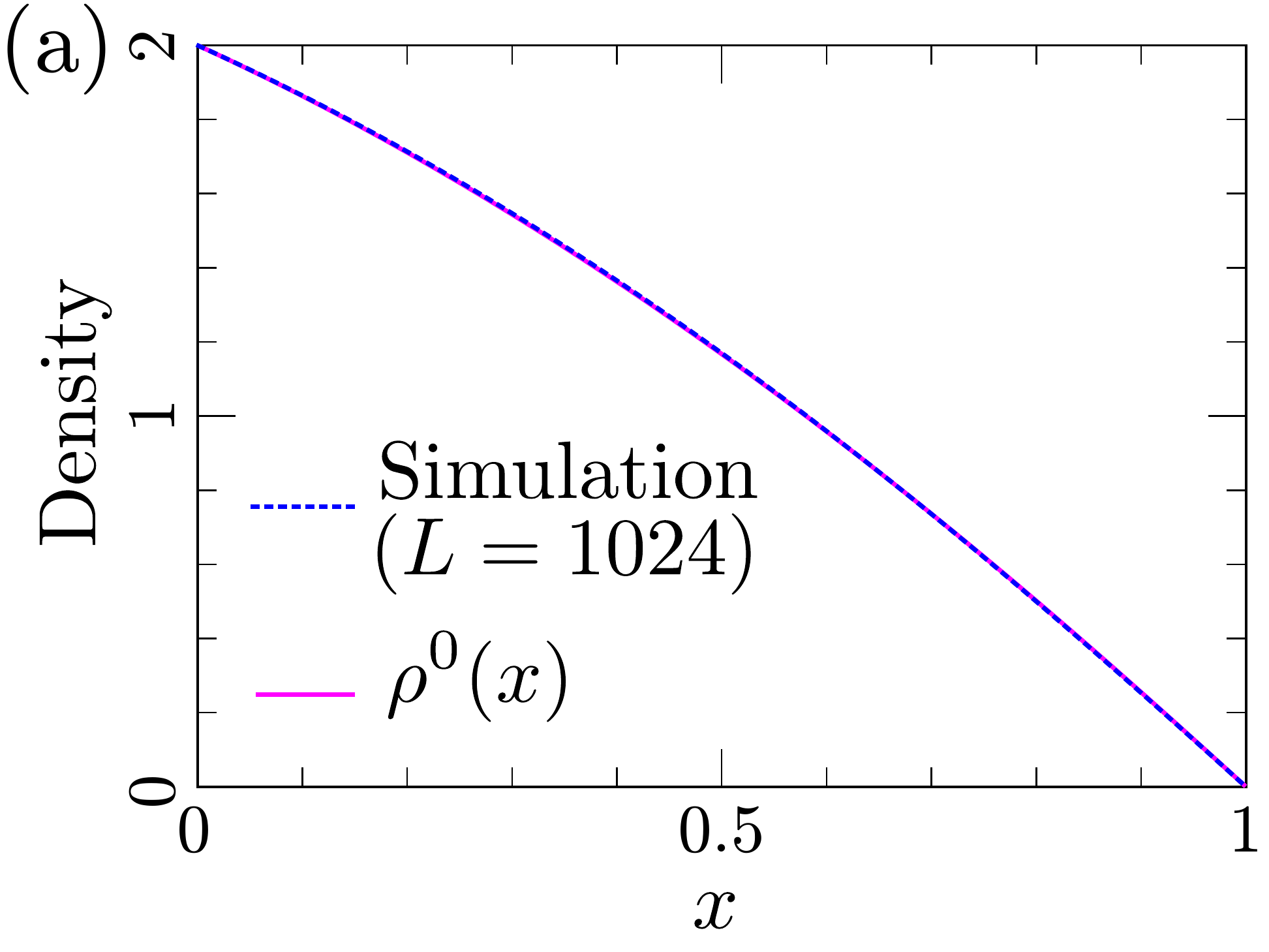}
 \includegraphics[width=60mm]{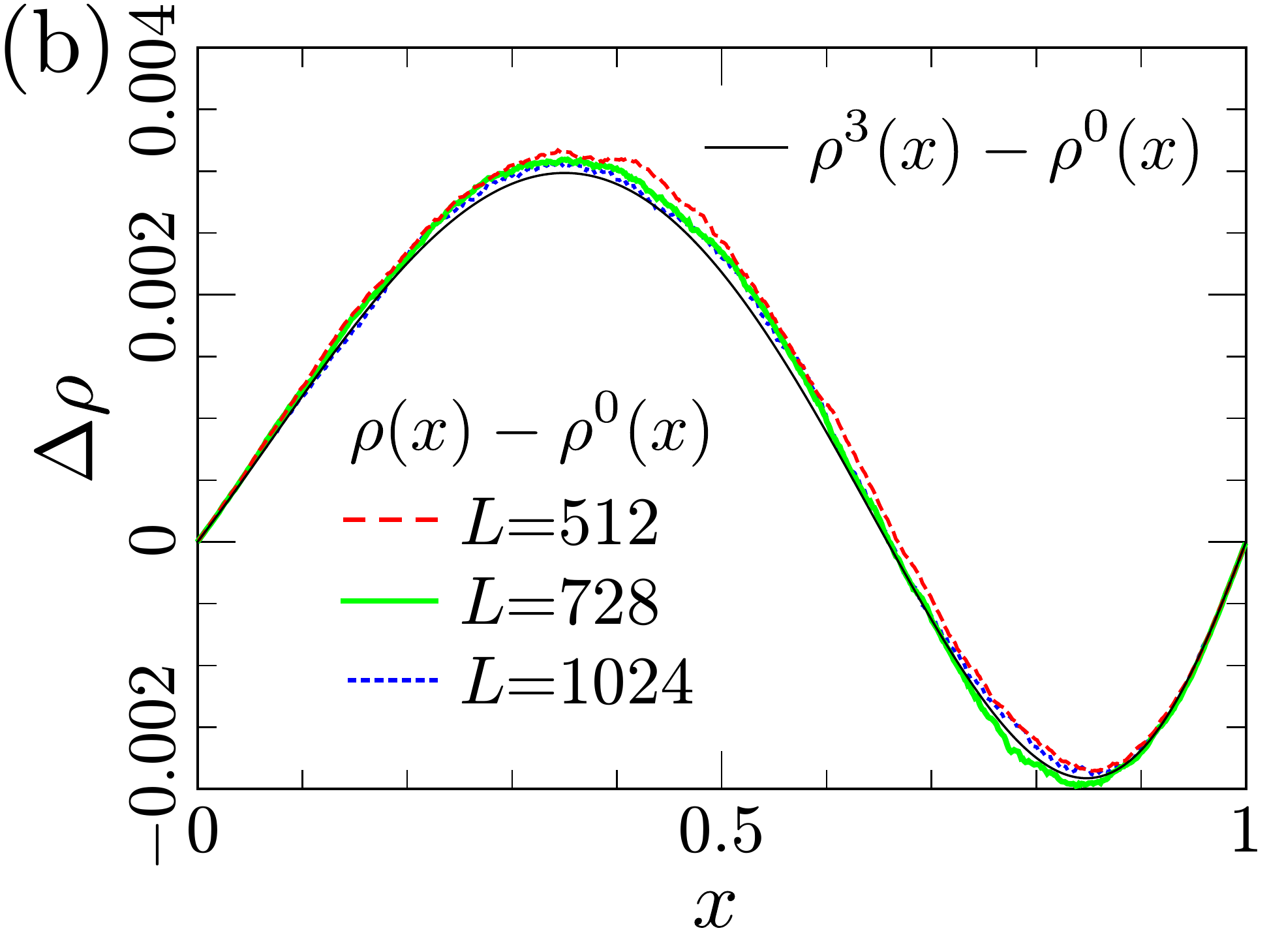}
 \includegraphics[width=60mm]{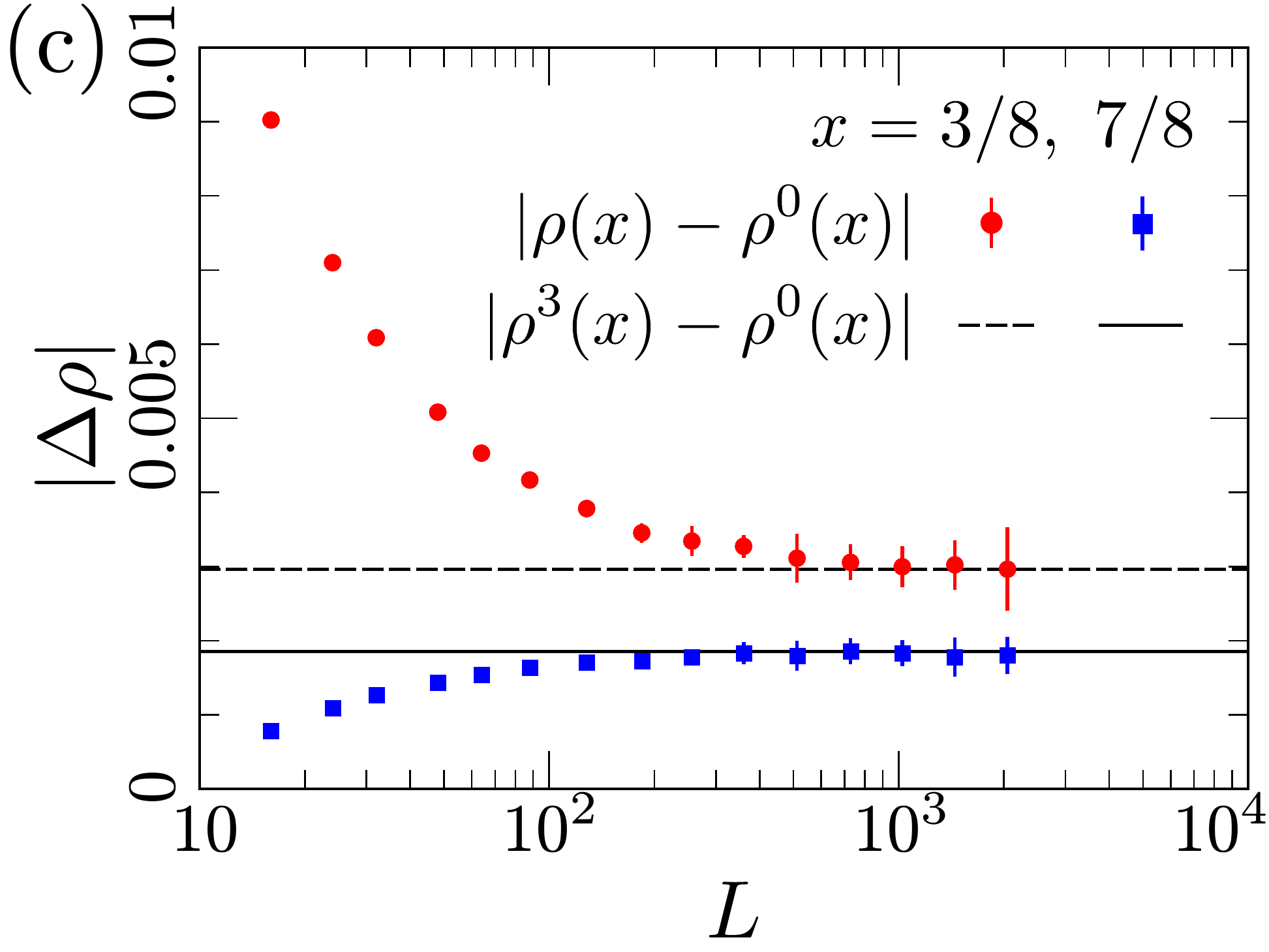}
 \caption{(a) $ \rho(x)$ for the system of size $ L=1024 $ and $ \rho^0(x) $
 obtained from Eq.~\eqref{eq:intD=xintD} with replacement $D\to D_0$. 
 (b) The difference between $ \rho^0(x) $ and $ \rho(x) $  observed in simulations. The difference $\rho^0 (x) - \rho^3 (x)$ is shown for comparison. 
  (c) The difference $\rho^0 (x) - \rho(x) $ for $ x=3/8 $ and $ 7/8 $, as a function of  system size. The difference $\rho^0 (x) - \rho^3 (x) $, for the same two values of $x$, are also shown.   For the three panels, we choose the boundary densities  $ (\rho_0,\rho_L) = ( 2,0 )$.  
 \label{fig:density}}
\end{center}
\end{figure}

In the long-time limit, the density profile in the open system becomes stationary and the current is uniform. The knowledge of $ D (\rho )$ allows us to determine the density profile by solving the stationary diffusion equation $\frac{d}{dx}\big( D (\rho )\frac{d\rho}{dx}\big) = 0$ and by imposing the boundary conditions matching the densities of reservoirs, $\rho(0) = \rho_0$ and $\rho(1) = \rho_L$. (Here, $x= i/L$ is the rescaled length.) This gives the implicit form of the density profile 
\begin{align}
\label{eq:intD=xintD}
 \int_{\rho_0}^{\rho(x)} D (\rho') d \rho' = x \int_{\rho_0}^{\rho_L} D (\rho') d \rho' , 
\end{align}
valid for $ 0\le x \le 1$. 
Since this form contains both $ \rho_0$ and $\rho_L$, the density profile $ \rho(x) $ depends, of course, on these boundary densities. Replacing the exact diffusivity $D(\rho)$ by $D_n(\rho)$ (for $n=1,2,3,\ldots$) derived in the previous section, leads to increasingly accurate approximations $ \rho^n(x) $ of the true profile $\rho(x)$. As shown in Fig.~\ref{fig:density} (a), the difference between $ \rho(x)$ and $\rho^0 (x) $ is already too small to be visible when $L=1024$, namely it is less than $ 0.004 $ in absolute value, see Fig.~\ref{fig:density} (b). However, the discrepancy between $ \rho(x)$ and $\rho^0 (x) $ is systematic and is not due to statistical errors: Fig.~\ref{fig:density} (c) indicates that it does not vanish in the $ L \to \infty $ limit. Figure \ref{fig:density} (b, c) also demonstrates that $ \rho(x) - \rho^0 (x) $ is well fitted by $\rho^3 (x)-\rho^{0} (x)$. Generally, $ \rho^n (x) $ with increasing $n$ provide more and more accurate predictions for $ \rho(x)$. 

If the gradient condition is satisfied, as for the misanthrope process, the exact bulk diffusivity is given by $D_0(\rho)$ and therefore $\rho^{0} (x)$ becomes identical to $ \rho(x)$ in the hydrodynamic limit $ L\to \infty $.

\subsection{Finite-size corrections to the product measure}

\begin{figure}
\begin{center}
 \includegraphics[width=60mm]{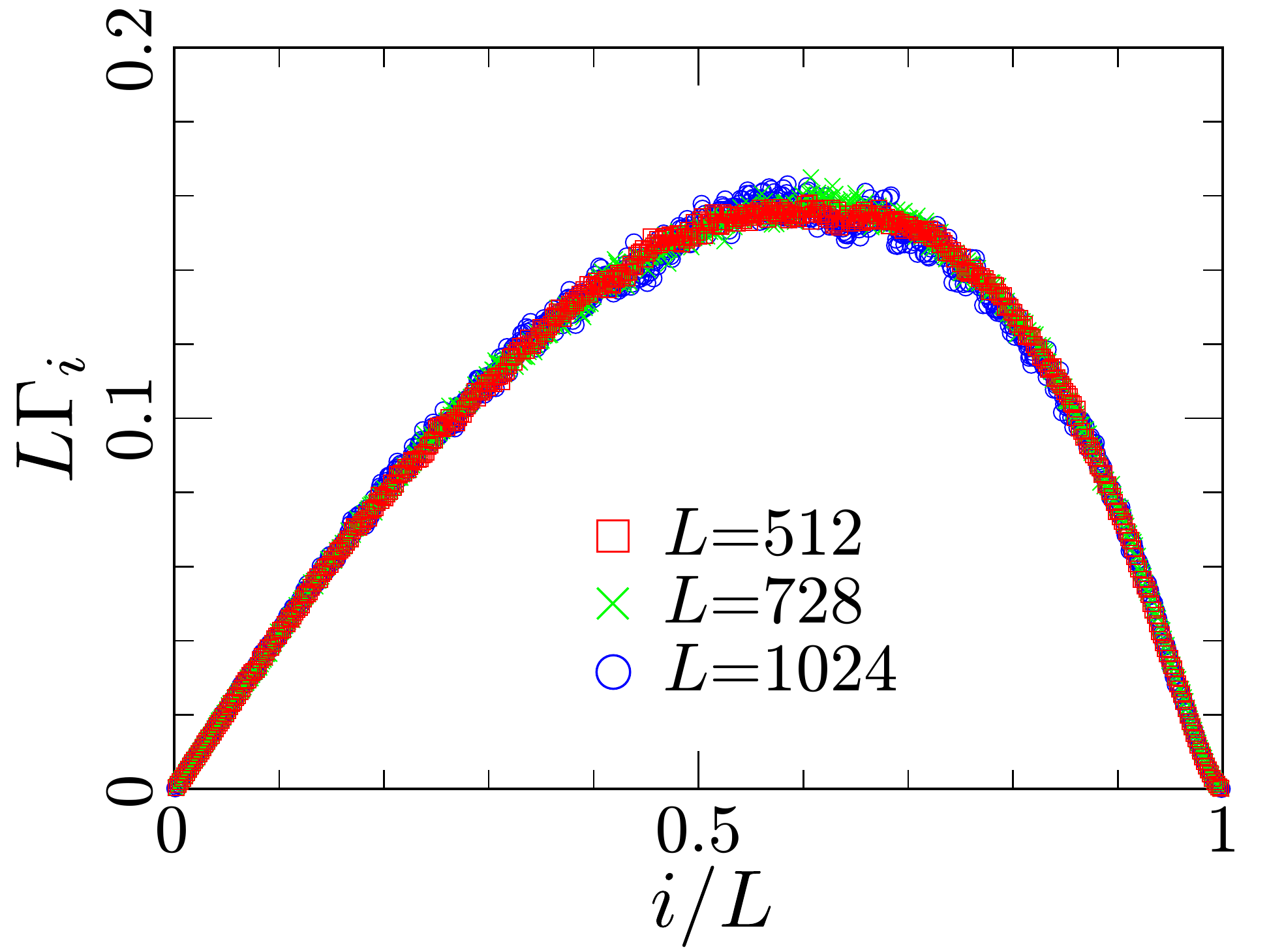}
 \caption{  The scaled correlation function $L\Gamma_i $ vs the scaled coordinate $i/L$ for three 
 different system sizes. The data collapse supports the emergence of the scaling expressed by 
 \eqref{eq:Qi=omega/L+o}.
 \label{fig:Gamma}}
\end{center}
\end{figure}

Because the stationary measure is not factorized in finite systems with open boundaries, the knowledge of the local densities 
\begin{align}
\label{eq:rhoi=} 
 \rho_i = \langle \tau_i \rangle = \mathbb P [ \tau_i = 1 ] + 2 \mathbb P [ \tau_i = 2 ] 
\end{align} 
is not sufficient to determine the stationary state. 
In an open system of size $L$, deviations from the product measure scale as $L^{-1}$ in the leading order. In order to probe these deviations, we first consider the quantity 
\begin{align}
\label{eq:Gammai=aP1P1-P0P2}
 \Gamma_i = a \mathbb P [ \tau_i = 1 ] ^2 - \mathbb P [ \tau_i = 0 ]\, \mathbb P [ \tau_i = 2 ] 
\end{align}
which vanishes for infinite systems and for finite systems on the ring, see Eq.~\eqref{eq:aX1X1-X0X2=0}. 
Figure~\ref{fig:Gamma} shows simulation results in the system with particle-uniform rates \eqref{eq:rates} and reservoir densities $ \rho_0 = 2 $ and $ \rho_L = 0 $. The plots indicate that $ \Gamma_i $ acquires a scaling form:
\begin{align} 
\label{eq:Qi=omega/L+o} 
\Gamma_i \simeq L^{-1} \omega(x) 
\end{align}
in the scaling limit
\begin{equation} 
\label{scaling:x} 
L\to\infty, \quad i\to\infty, \quad x=\frac{i}{L}=\text{finite}. 
\end{equation}

Next we consider the one-point functions $\mathbb P [ \tau_i = r ]$. The corresponding finite-size corrections have the form similar to \eqref{eq:Qi=omega/L+o}:
\begin{align} 
\label{eq:P=X+kappa/L}
 \mathbb P [ \tau_i = r ] - W_r ( \rho_i ) \simeq L^{-1} \kappa _r(x)
\end{align}
with scaling functions $ \kappa_r$ obeying 
\begin{align}
 \kappa_0 = - \frac{\kappa_1}{2} =\kappa_2= 
- \frac{ \omega (x) }{ \sqrt{ 1 - (1-4a) \rho(x) [ 2-\rho(x) ] } } \, . 
 \end{align}
We verified \eqref{eq:P=X+kappa/L} for the particle-uniform rates \eqref{eq:rates} and expect \eqref{eq:P=X+kappa/L} to hold for a generic choice of hopping rates. 

\begin{figure}
\begin{center}
 \includegraphics[width=35mm]{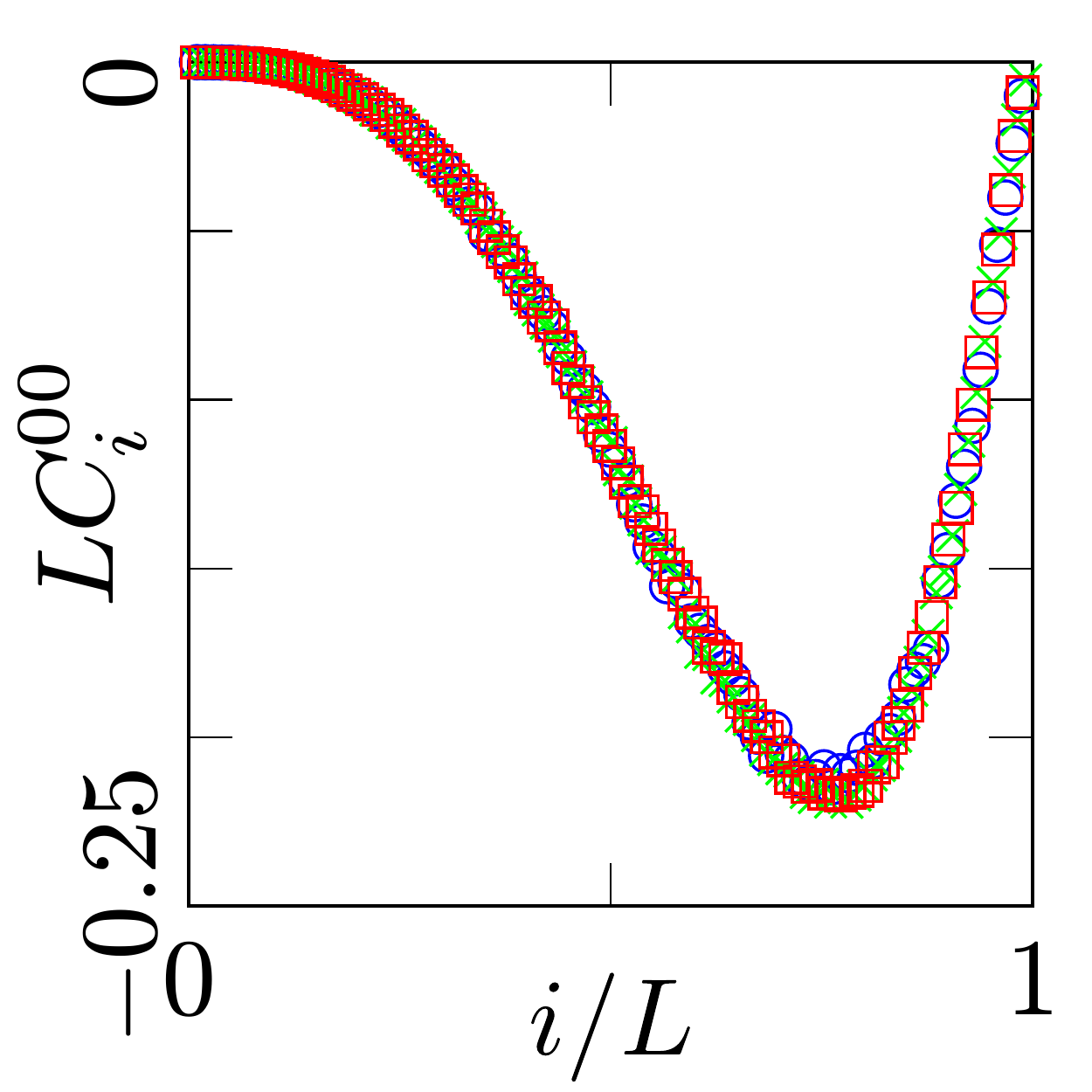}
 \includegraphics[width=35mm]{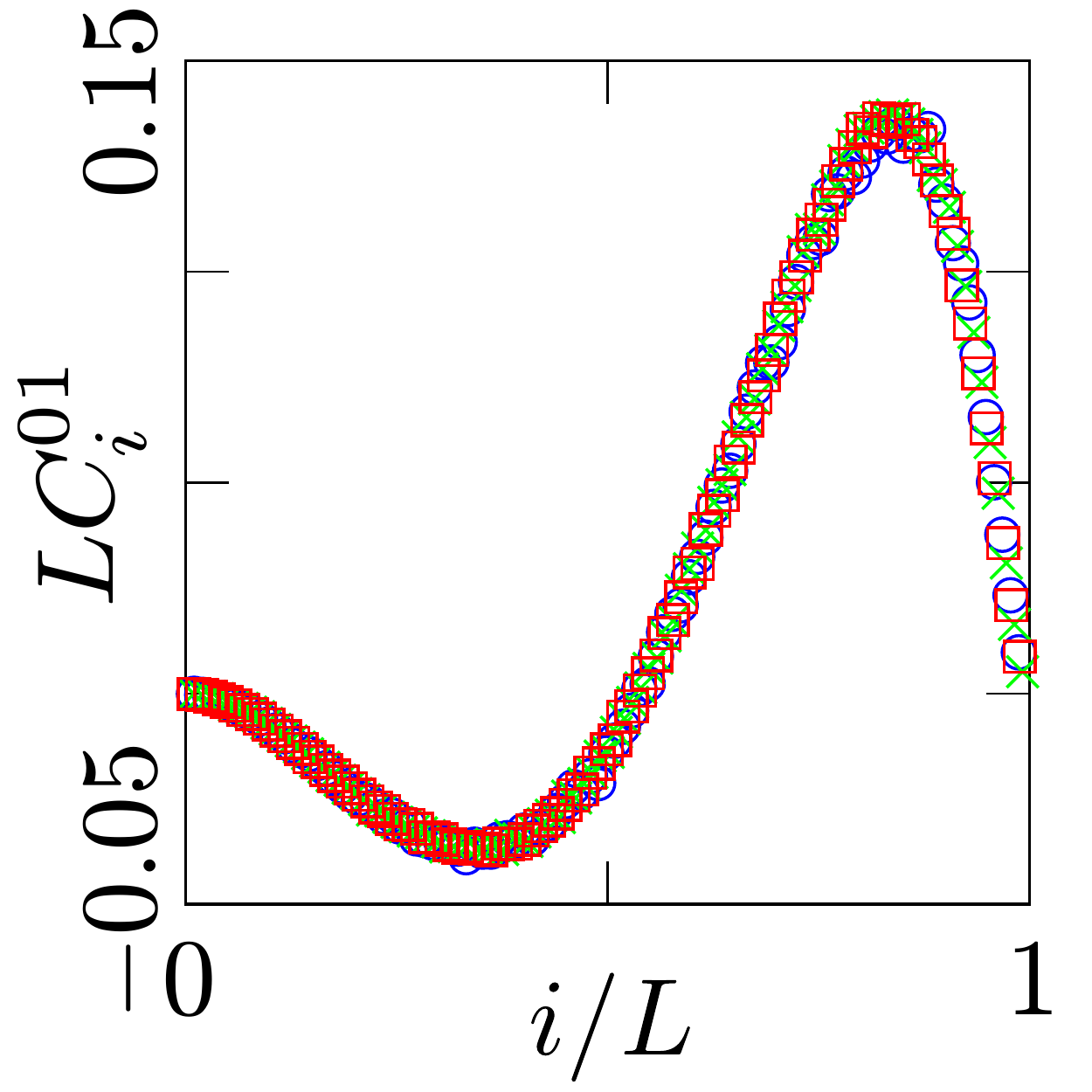}
 \includegraphics[width=35mm]{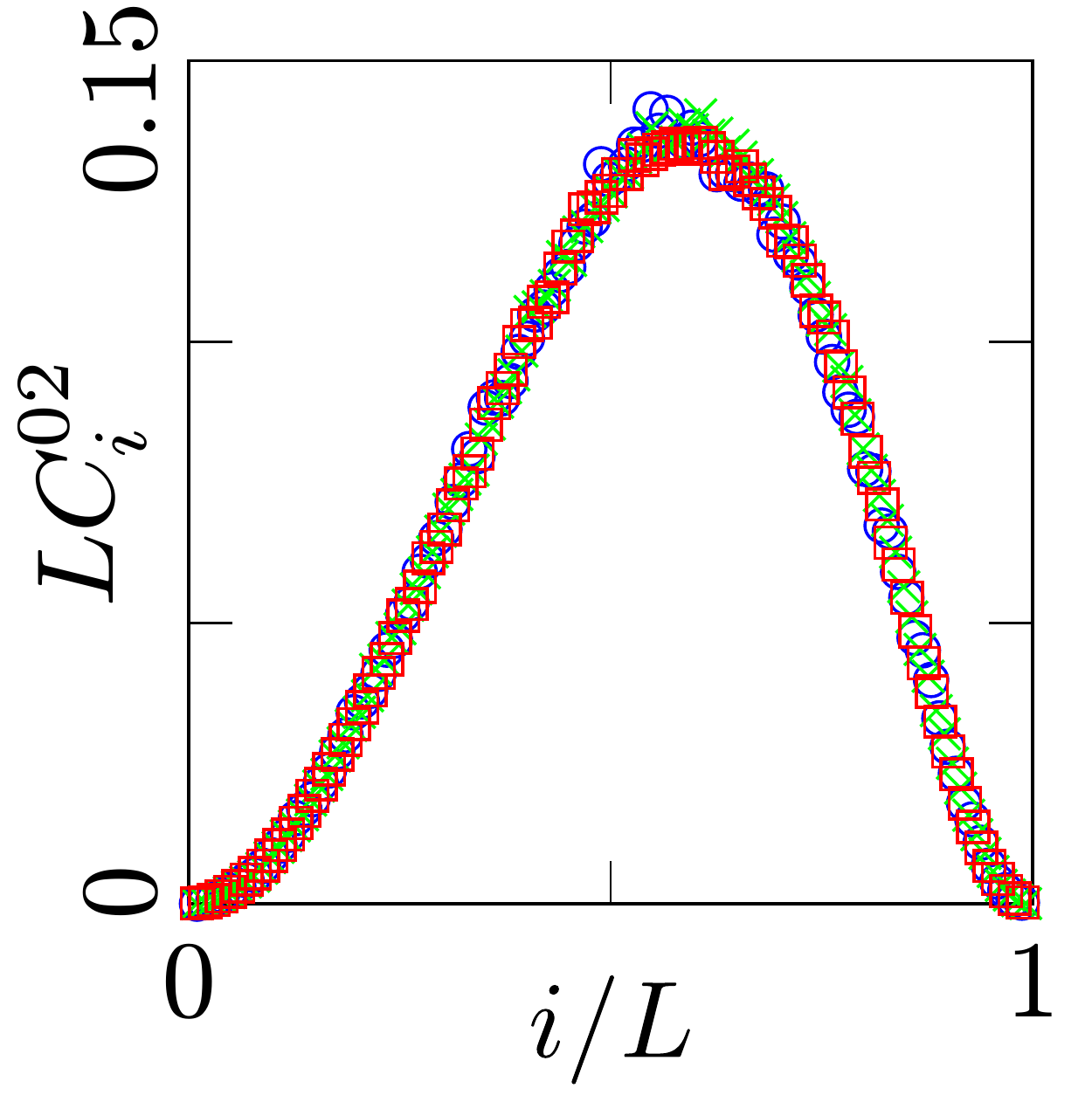}
 \includegraphics[width=35mm]{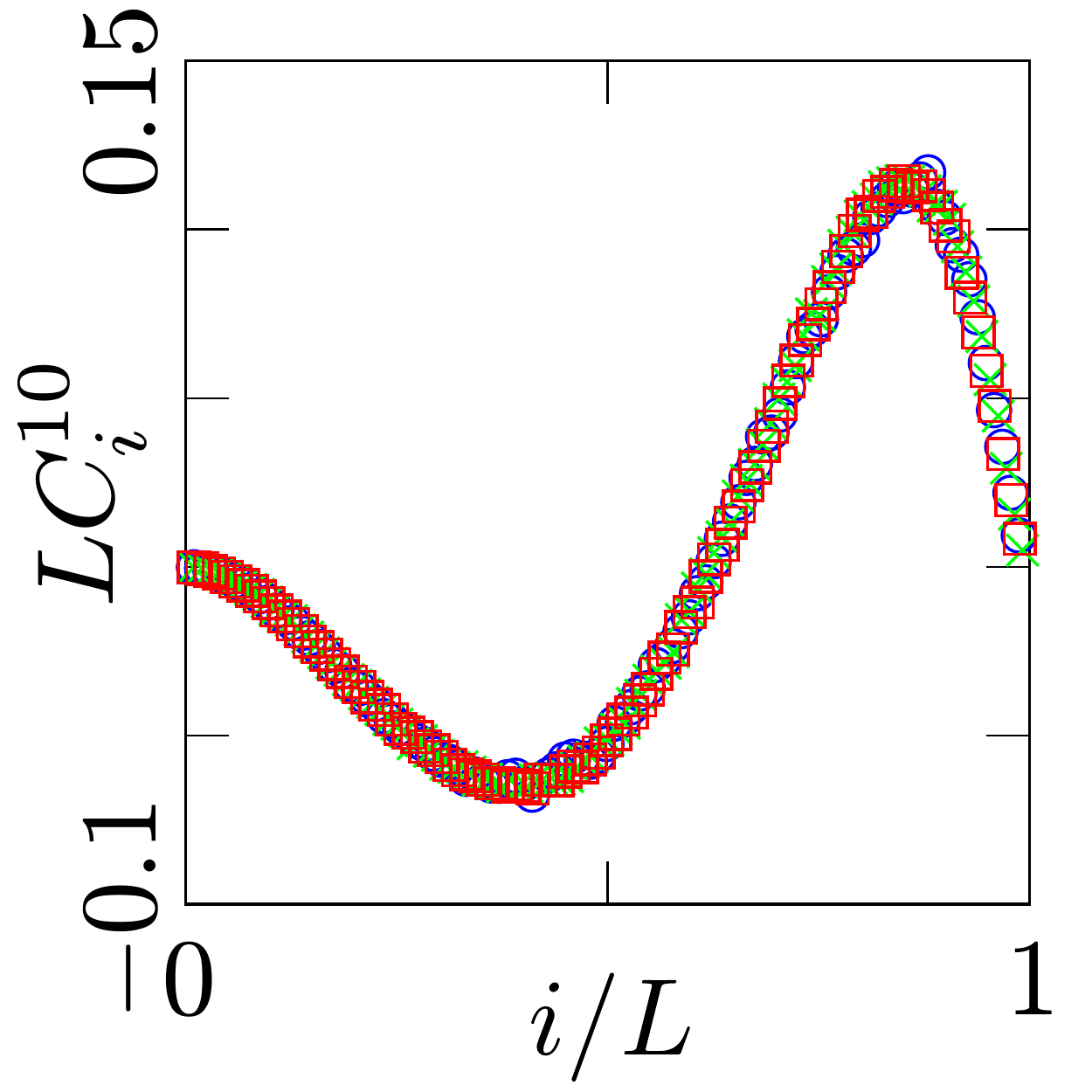}
 \includegraphics[width=35mm]{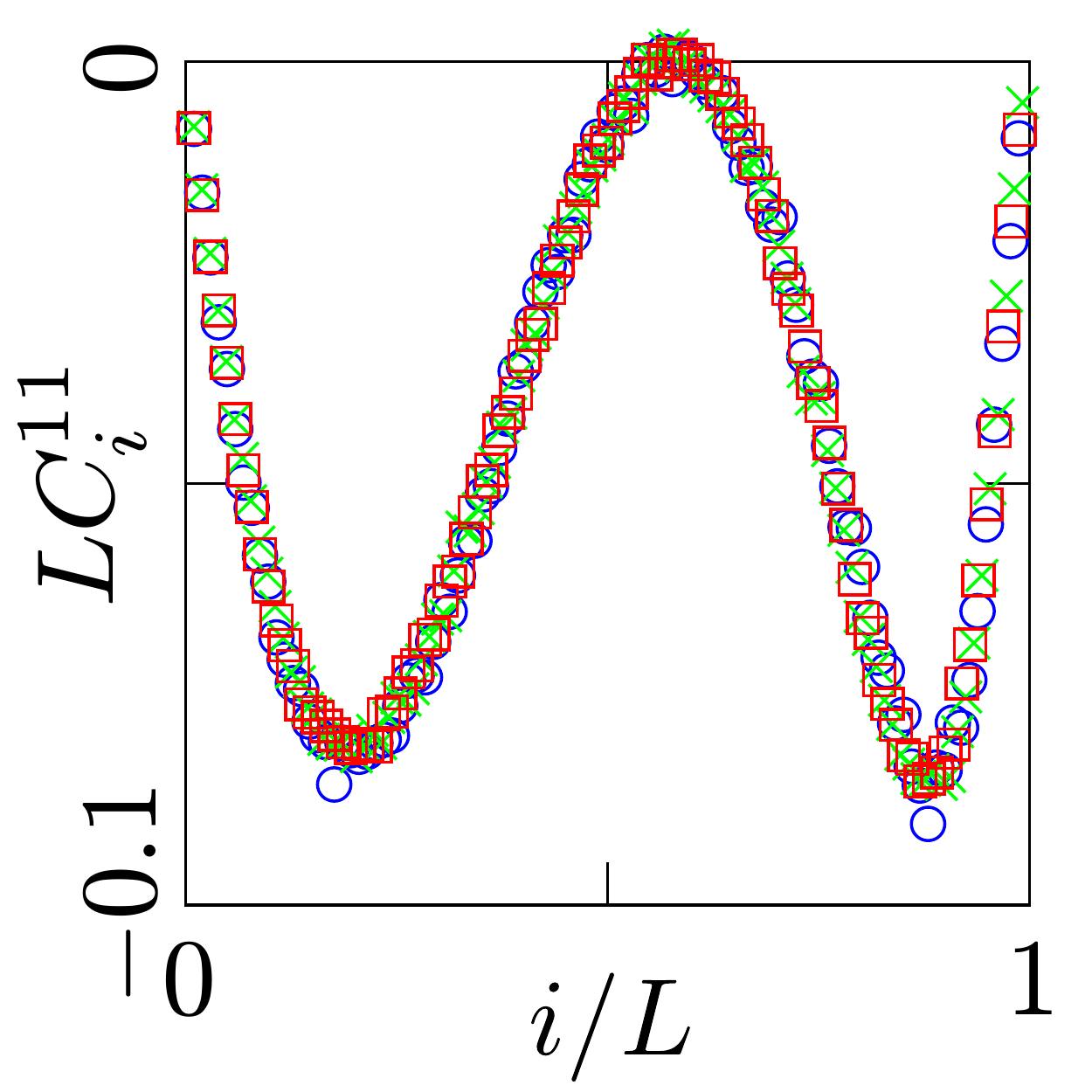}
 \includegraphics[width=35mm]{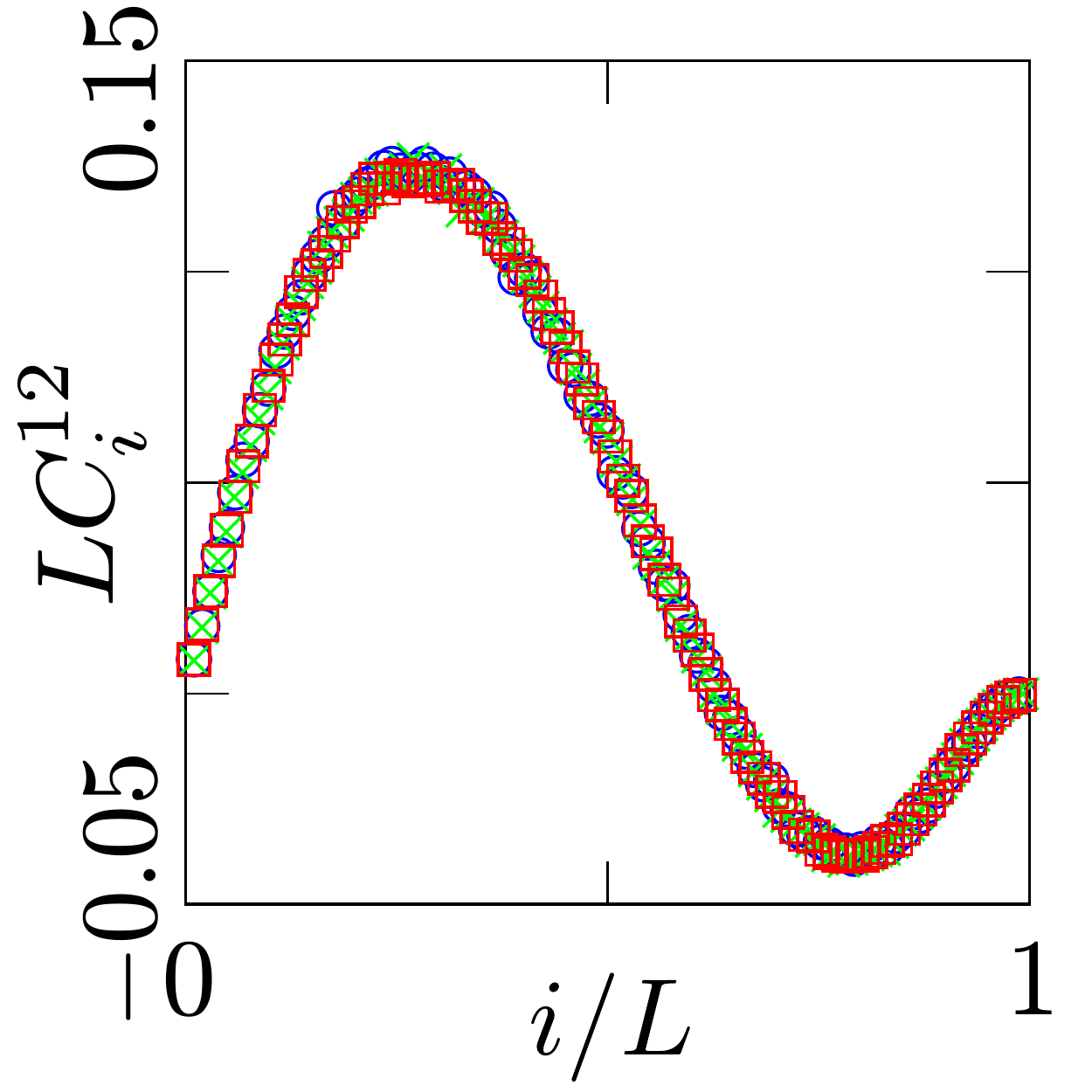}
 \includegraphics[width=35mm]{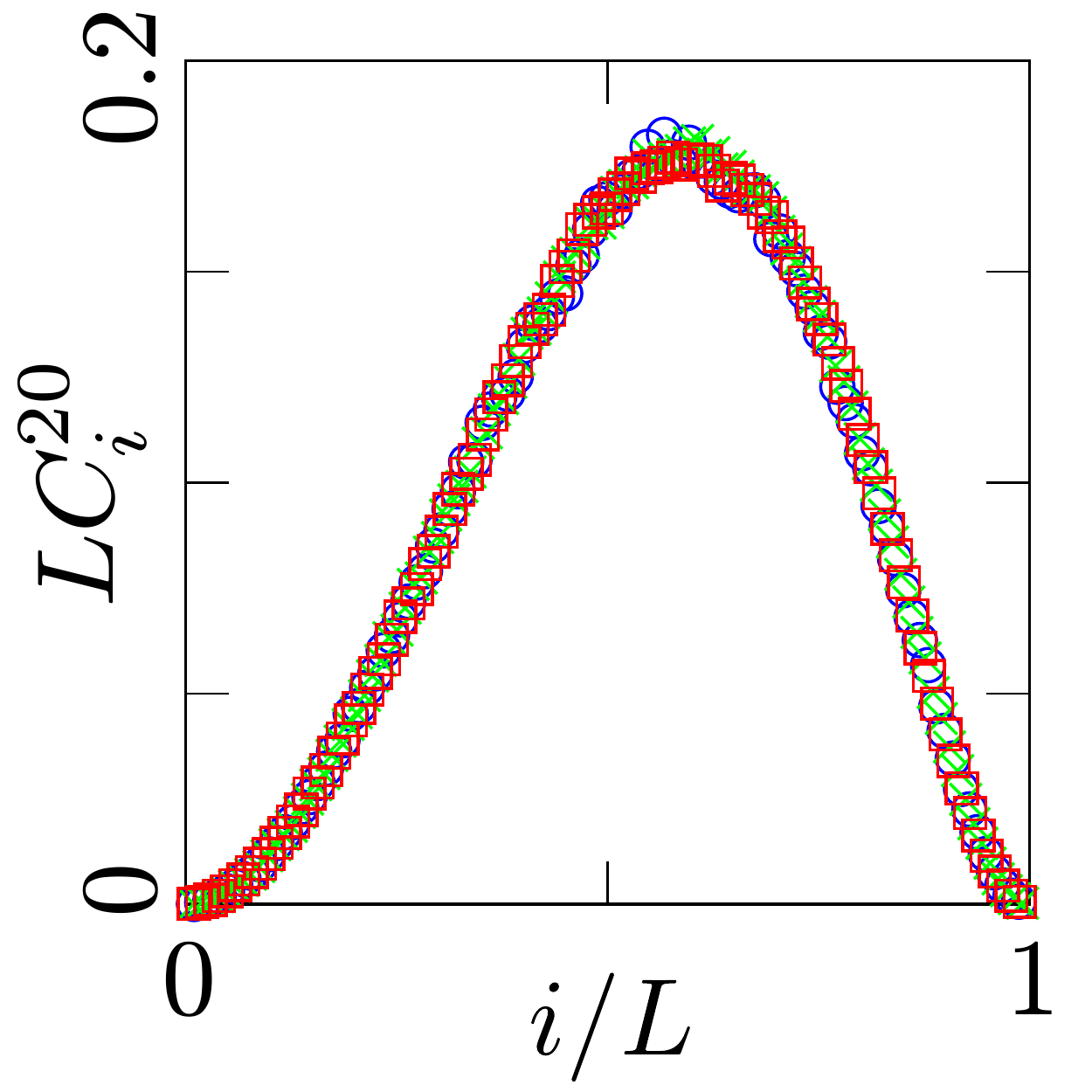}
 \includegraphics[width=35mm]{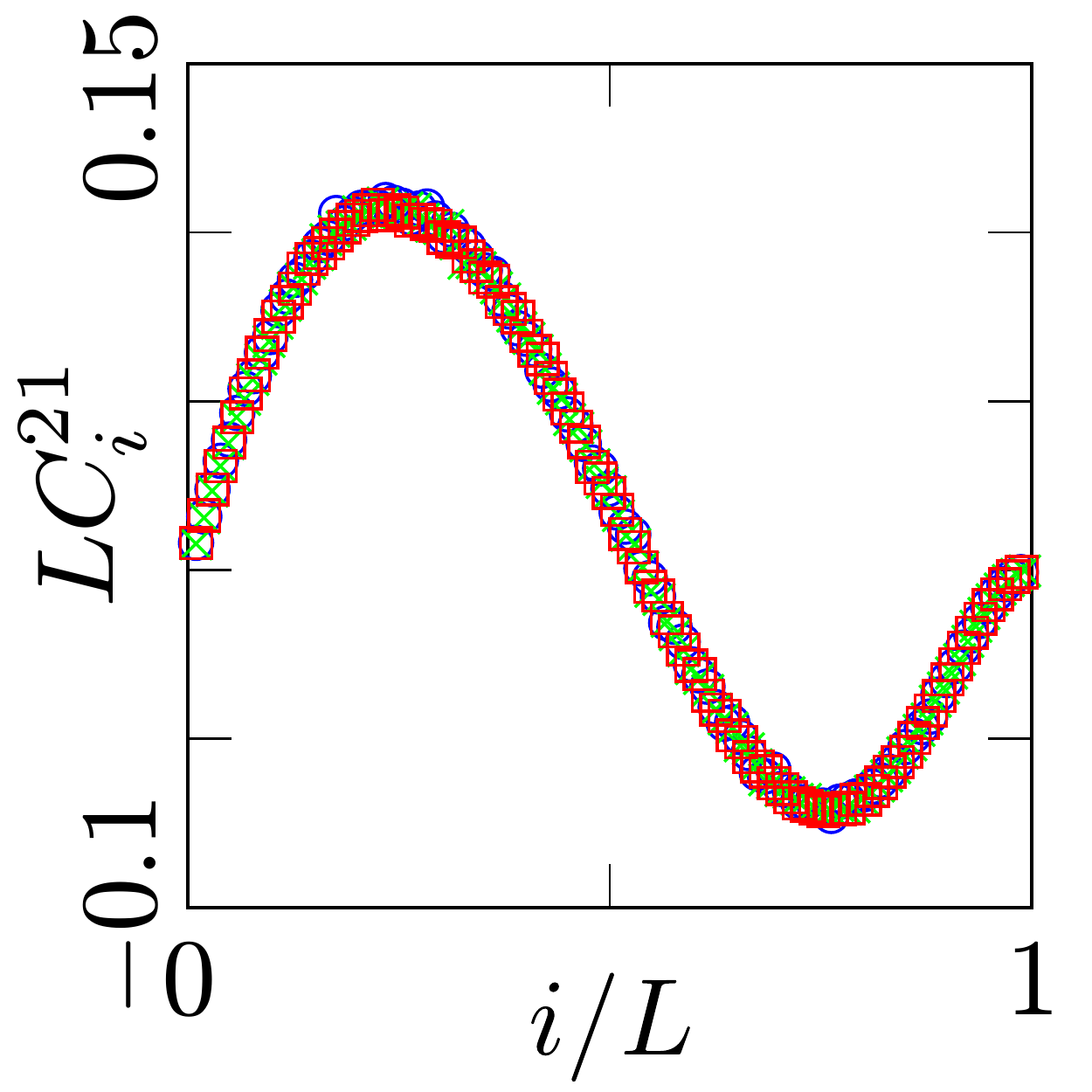}
 \includegraphics[width=35mm]{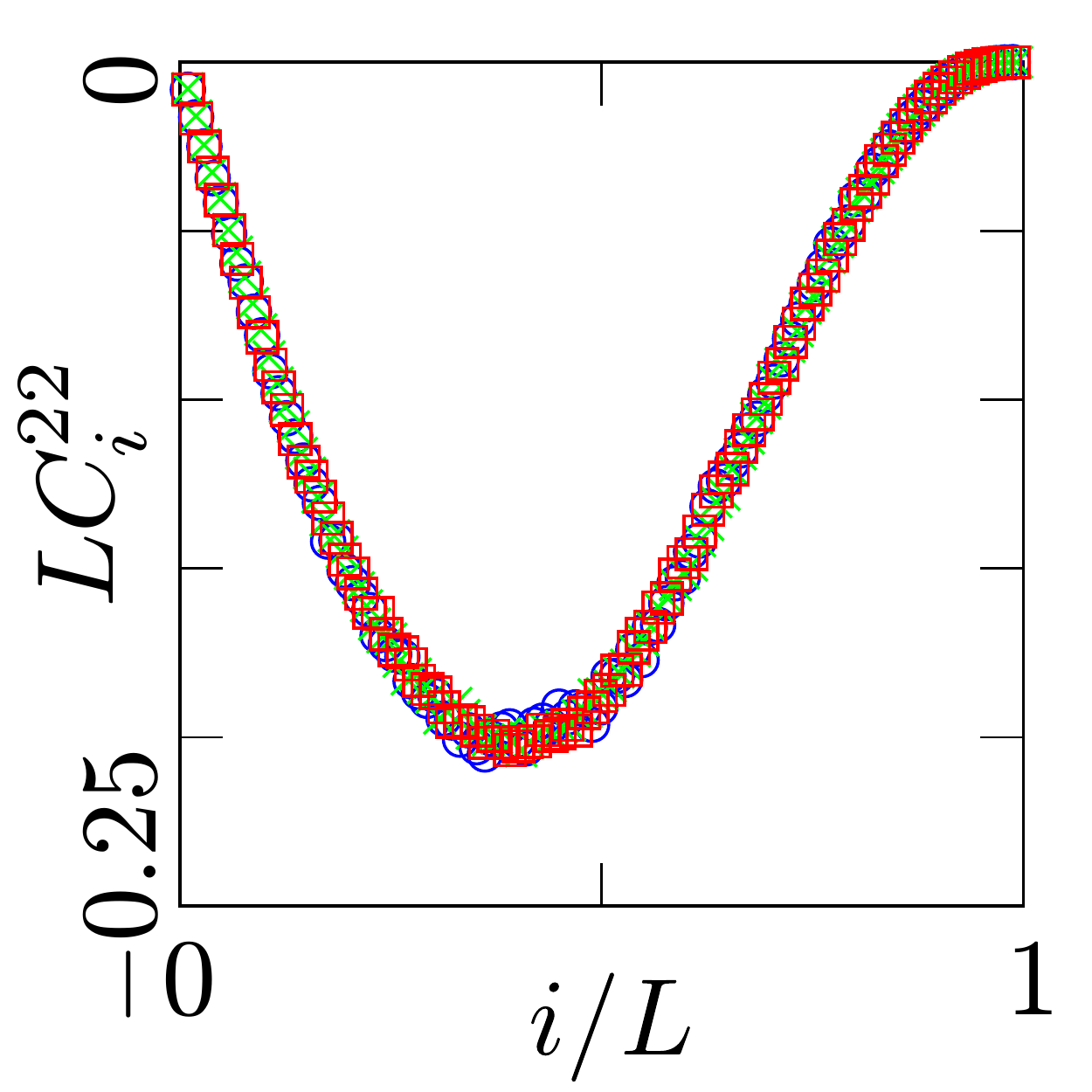}
 \includegraphics[width=35mm]{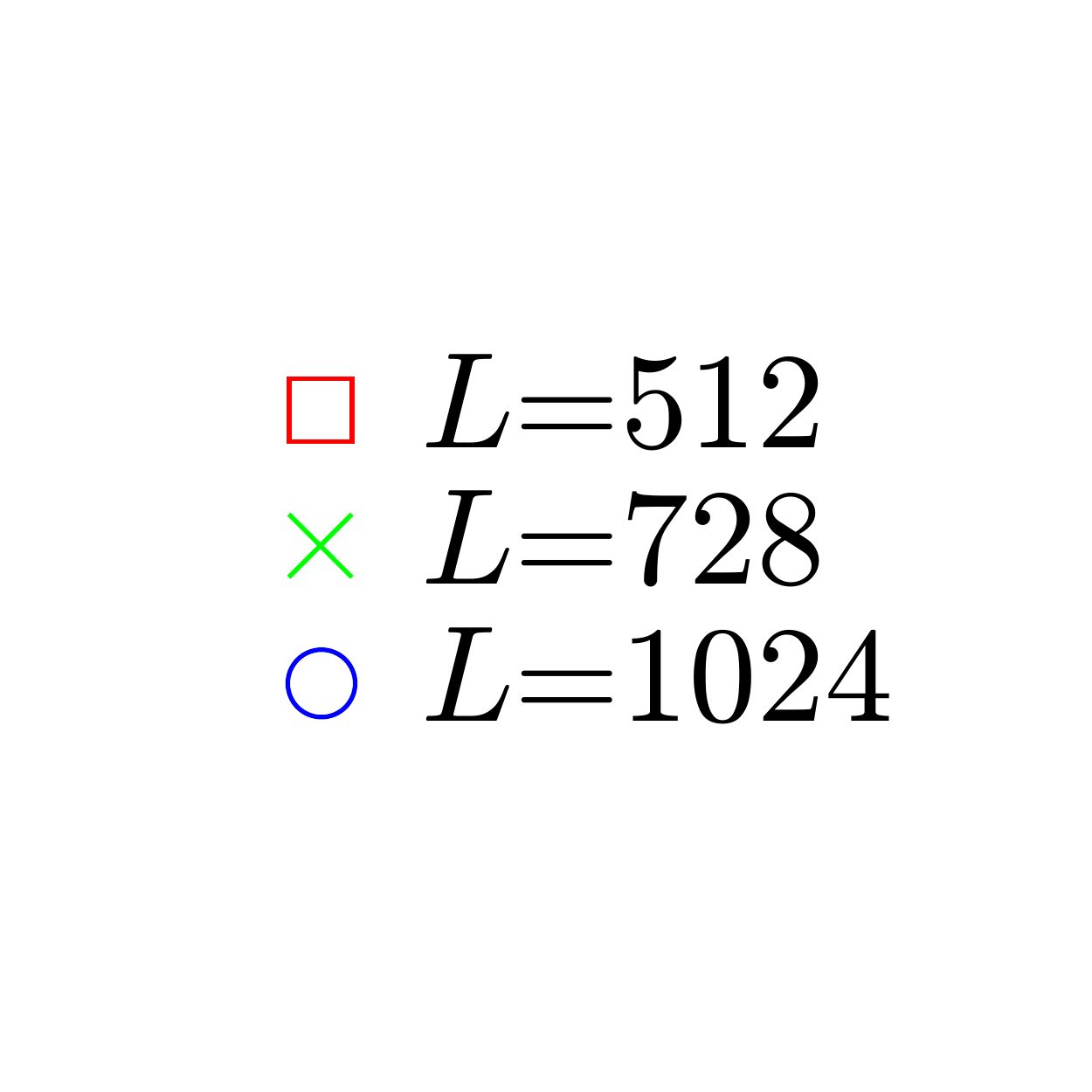}
 \caption{  Correlation functions \eqref{eq:Cmni} multiplied by the system size $L$. The data collapse
 supports the emergence of the scaling behavior 
\eqref{eq:C=f/L}. The sum rules \eqref{eq:C+C+C=0(r)}--\eqref{eq:C+C+C=0(s)} are obeyed. } 
 \label{fig:correlations}
\end{center}
\end{figure}

The pair correlation function between two adjacent sites plays a crucial role in the following analysis. We use notations
\begin{subequations}
\begin{align}
\label{def:Xi}
X_i^{rs} &= \P [ \tau_i = r \wedge \tau_{i+1} = s ]\ \\
\label{def:Y}
Y_i^{rs} &= \P [ \tau_i = r ]\,\P [ \tau_{i+1} = s ]
\end{align}
\end{subequations}
with $ r,s\in \{0,1,2\} $ and focus on the connected version of the pair correlation function
\begin{equation}
 \label{eq:Cmni}
C_i^{rs} = X_i^{rs} - Y_i^{rs}\,.
\end{equation}
For a periodic or infinite system, $ C_i^{rs} = 0 $. However, this is generally not true for open systems of size $L$. In the scaling limit \eqref{scaling:x}, the pair correlation function acquires a scaling form
\begin{align} 
\label{eq:C=f/L}
 C_i^{rs} \simeq L^{-1} \varphi^{rs}(x) \,.
\end{align}
Note that these correlation functions satisfy simple sum rules 
\begin{subequations}
\begin{align}
\label{eq:C+C+C=0(r)}
 \varphi^{r0}(x) + \varphi^{r1 }(x)+ \varphi^{r2 }(x) = 0& \quad ( r=0,1,2 ) \\
 \varphi^{0s}(x) + \varphi^{1s}(x)+ \varphi^{2s}(x) = 0& \quad ( s=0,1,2 ) 
\label{eq:C+C+C=0(s)}
\end{align}
\end{subequations}
obtained by summing over the possible occupations of the right-hand side of  \eqref{eq:Cmni}. Only four of the nine functions $\varphi^{rs}$ are independent. In Fig.~\ref{fig:correlations}, all nine functions are plotted. The sum rules are obeyed thereby providing a check of our simulations.

\subsection{Hydrodynamic limit for the current}
\label{sec:current}

We finally derive an alternative formula for the diffusion coefficient $D(\rho)$ by taking the continuous limit of the exact expression for the microscopic current. It is crucial to keep the dominant corrections to the product measure defined in \eqref{eq:C=f/L}. This approach allows us to identify the missing contribution in the naive hydrodynamic limit of \cite{bib:AKM}. The alternative formula for $D(\rho)$ is again checked by numerical simulations. 

The current between site $i$ and $i+1 $ reads
\begin{align}
\label{eq:Ji=}
 J_i = \sum_{ 1\le r \le2 \atop 0\le s \le 1 } p_{rs} ( X_i^{rs} - X_i^{sr} ) 
\end{align}
and it can be re-written as 
\begin{subequations}
\begin{align}
\label{eq:Ji=hatJi+(ppp)(CC)} 
&J_i = \widehat J_i+ (p_{21}+p_{10}-p_{20})( C_i^{21} - C_i^{12} )\,, \\
&\widehat J_i = \sum_{ 1\le r \le2 \atop 0\le s \le 1 } p_{rs} ( Y_i^{rs} - Y_i^{sr} ), 
\label{eq:hatJi=} 
\end{align}
\end{subequations}
see Appendix \ref{sec:app-currents} for details.
The term $ \widehat J_i $ involves only the products $Y_i^{rs}$ of one-point functions. The second 
term on the right-hand side of \eqref{eq:Ji=hatJi+(ppp)(CC)} vanishes when 
the 2-GEP satisfies the gradient condition (i.e., for the misanthrope process). 

In the scaling limit \eqref{scaling:x}, the term $ \widehat J_i $ becomes 
\begin{equation}
\label{grad-J}
 \widehat{J}(x) = - \frac{1}{L} D_0 ( \rho ) \frac{d\rho}{dx}\, ,
\end{equation}
where $ D_0 $ is the zeroth order approximation of the diffusivity (see Appendix \ref{sec:limJhat} for details). 
Note that \eqref{grad-J} is identical to the naive hydrodynamic limit \cite{bib:AKM}. 

The second term on the right-hand side of \eqref{eq:Ji=hatJi+(ppp)(CC)} also simplifies in the scaling limit \eqref{scaling:x}, namely it turns into $L^{-1}\mu(x)$ with 
\begin{align}
\label{eq:mu=phi01-phi10} 
\mu (x) = (p_{21}+p_{10}-p_{20}) \left[ \varphi^{21}( x )- \varphi^{12} ( x ) \right].
\end{align} 
Combining \eqref{grad-J} and \eqref{eq:mu=phi01-phi10} we arrive at 
\begin{align} 
 L J(x) \simeq - D_0( \rho ) \frac{d\rho}{dx} + \mu (x) ,
\label{Hydro-Current}
\end{align}
which in conjunction with Fick's law implies that
\begin{align}
\label{eq:D=D0-mudx/drho}
 D(\rho) = D_0 (\rho) - \mu ( x ) \frac{ d x}{d\rho }\, . 
\end{align}
Here $ x=x (\rho ) $ is the inverse function of the stationary density profile $ \rho(x) $ given by \eqref{eq:intD=xintD}. Equation \eqref{eq:D=D0-mudx/drho} allows us to obtain numerical plots of $D(\rho)$ since $ x=x (\rho ) $ and $ \mu ( x )$ can be determined accurately from simulations. (This procedure was used in preparing Fig.~\ref{fig:D0D2D3sim}.)

Rewriting Eq.~\eqref{eq:D=D0-mudx/drho} and using the implicit equation for the profile \eqref{eq:intD=xintD} we obtain 
\begin{align}
\label{eq:rhon}
 \mu (x) = \big[ D_0(\rho) - D (\rho) \big] \frac{d\rho}{dx} 
 = \left[ 1 - \frac{ D_0(\rho)} {D (\rho)} \right] I
\end{align}
with 
\begin{equation}
I \equiv - \int_{\rho_0}^{\rho_L} D (\rho) d \rho . 
\label{TotalCurrent}
\end{equation}

\begin{figure}
\begin{center}
 \includegraphics[width=60mm]{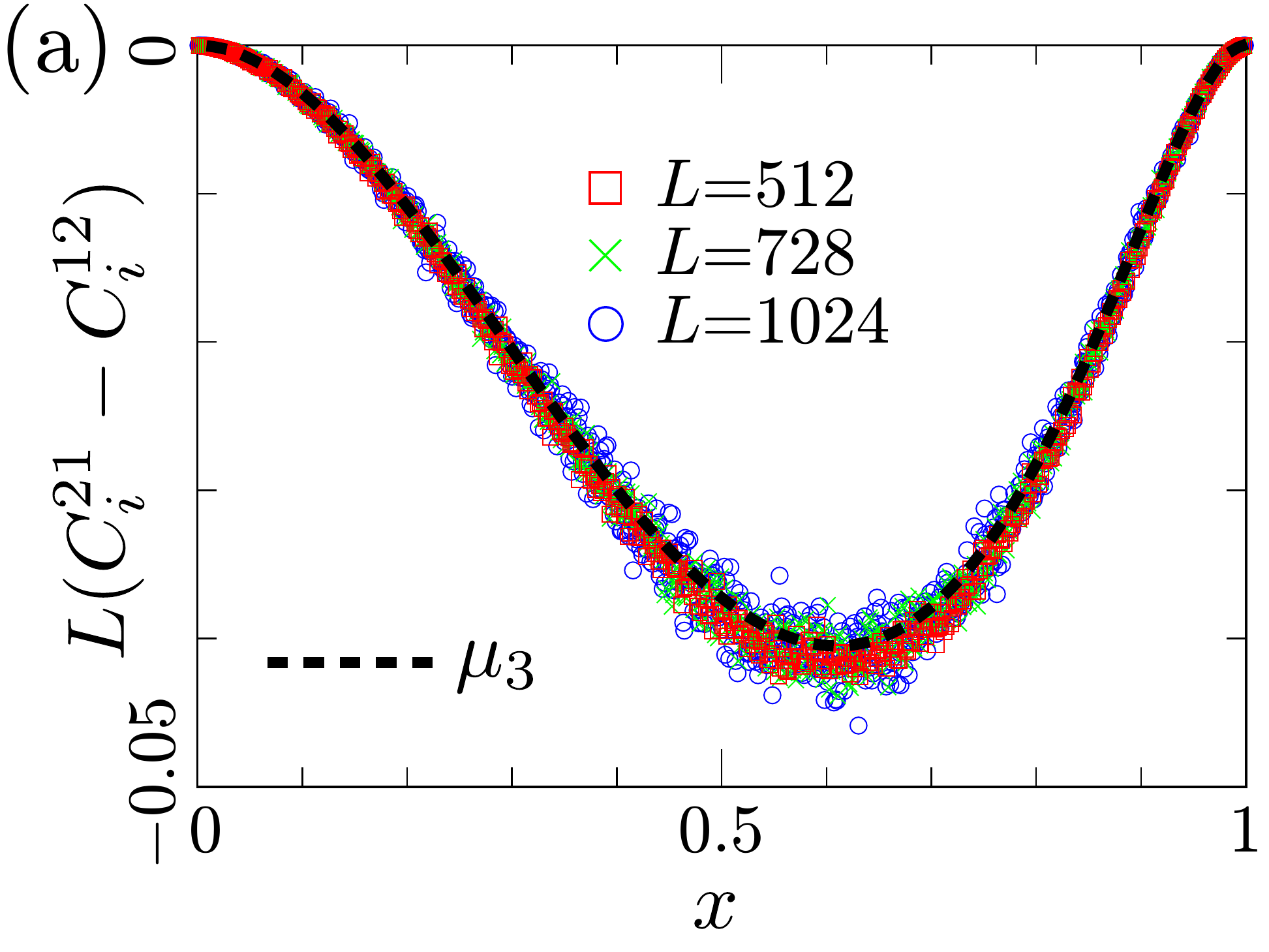}
 \includegraphics[width=60mm]{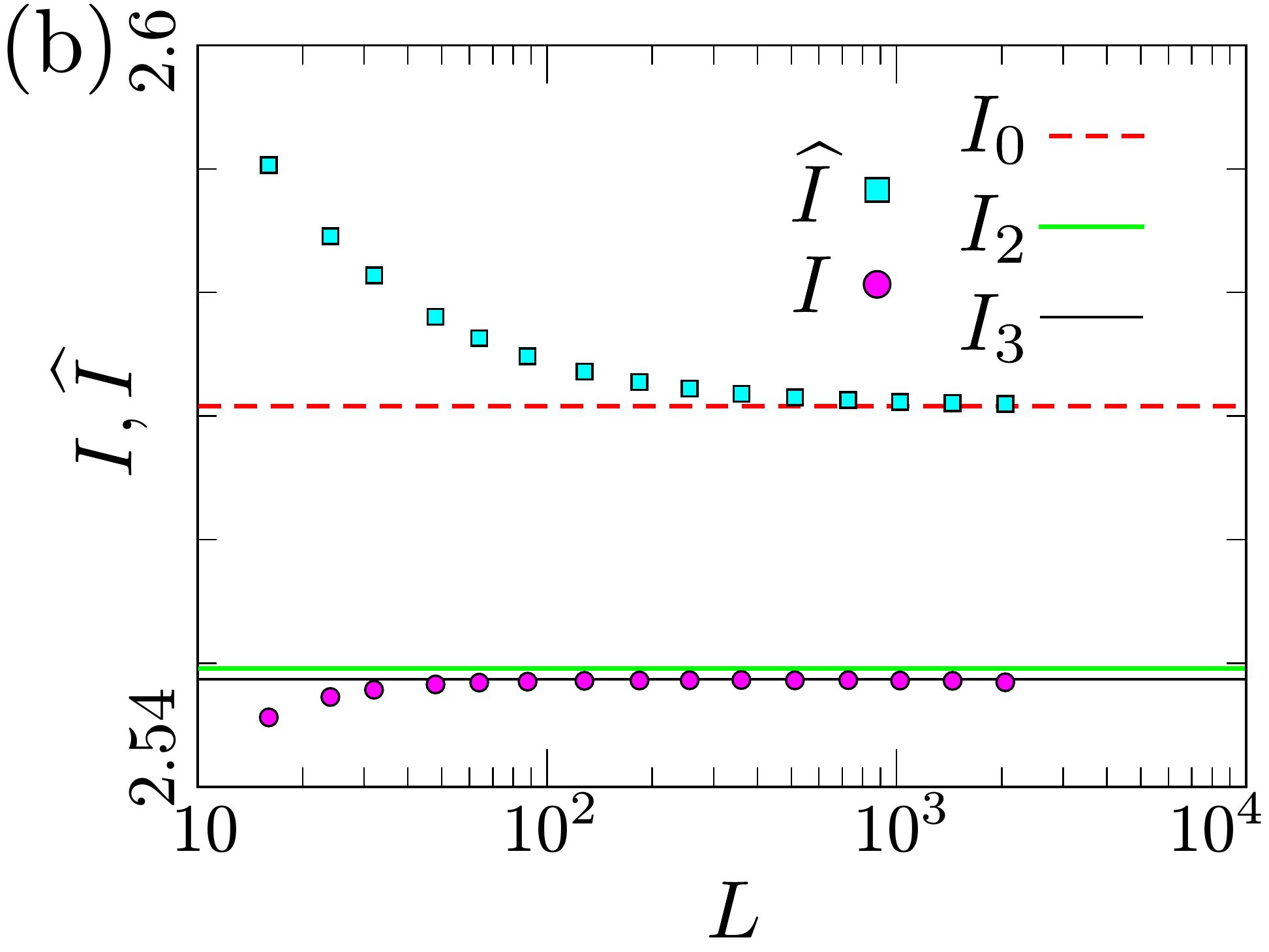}
 \caption{  (a) $L(C^{21}_i-C^{12}_i)$ vs $x=i/L$ for three different system sizes; 
 $ \mu_3$ [Eq.~\eqref{eq:mun}]  is shown for comparison. 
 (b) Simulation results for $ I = \int_0^2 D (\rho) d\rho$ and 
 $\widehat{I} = \int_0^2 D_0 (\rho) d\rho\, $. The estimates $I_n$ 
 with $n=0,2,3$ are also shown. 
 \label{fig:mu-currents}}
\end{center}
\end{figure}

Replacing $D$ by $D_n$, one obtains
\begin{equation}
\label{eq:mun}
 \mu_n(x) = \left[ 1 - \frac{ D_0 (\rho) }{ D_n(\rho) } \right] I_n, ~~ I_n = - \int_{\rho_0}^{\rho_L} D_n (\rho) d \rho\,.
\end{equation}
Simulation results for $\mu (x)$ are well fitted by $\mu_3(x)$, see Fig.~\ref{fig:mu-currents}. In Fig.~\ref{fig:mu-currents} we also plot $I$ defined in \eqref{TotalCurrent} as a function of $L$. We observe an excellent agreement between $ I $ (simulations) and $ I_3 $ (theory) as $ L $ increases. We also verify that $\widehat{I} \equiv - \int_{\rho_0}^{\rho_L} D_0 (\rho) d\rho\, $, converges to $ I_0 $ when $L \to \infty$. This confirms the usefulness of Eq.~\eqref{eq:rhon} for numerical measurements of the diffusion coefficient, as an alternative to the standard Green-Kubo formula.

\section{Conclusions}
\label{sec:conclusions}

Interacting many-particle systems generically exhibit macroscopic behaviors 
that can be described by fluctuating hydrodynamics. This is particularly well understood in 
the realm of stochastic lattice gases with symmetric hopping. 
In these diffusive lattice gases, the hydrodynamic behavior is described by the single scalar function, 
the coarse-grained density $\rho(x,t)$, satisfying the diffusion equation. 
Fluctuations and large deviations in diffusive lattice gases
are described by two coupled scalar fields evolving according to equations of the macroscopic
fluctuation theory. Remarkably, all details of the microscopic hopping rules
are encapsulated in the diffusion coefficient $D(\rho)$ and the conductivity $\sigma(\rho)$. 
These transport coefficients enter into the diffusion equation and the governing equations of the macroscopic
fluctuation theory. The determination of the transport coefficients is a challenge even for the simplest
diffusive lattice gases. Apart from very special models, known as gradient systems, 
for which a simple perturbation approach
provides the correct answer, there exists no general technique 
to derive closed formulas for the transport coefficients.

In this work, we used the Varadhan-Spohn variational formula for the diffusion coefficient and employed an approach resembling the Ritz method to derive analytical approximations for $D$ to an arbitrary degree of precision. 
The Varadhan-Spohn formula is essentially the Green-Kubo formula for diffusive lattice gases. 
We illustrated our approach by investigating the 2-GEP, a generalization of the symmetric 
exclusion process in which a site can accommodate at most two particles.
The 2-GEP is nongradient gas with unknown transport coefficients. 
For the 2-GEP in one dimension we showed that the Varadhan-Spohn 
variational formula can be used to derive analytical approximations 
to $D$ in a systematic manner---after a few iterations,
the precision of the order of a few parts per million is reached. Thus the 
simplest approximations are already remarkably accurate resembling the success
of continuous fraction approximations of irrational numbers. We emphasize that our 
approximate expressions for the diffusion coefficient are intrinsically
different from perturbative, Taylor-type, expansions around some special value of the density;
on the contrary, they are global and uniform estimates, valid on the whole range of admissible densities. 
In order to check for the practicality of our approximation procedure, we performed
simulations in open systems coupled to reservoirs
at different densities. The precise variational 
estimate of the diffusivity allowed us to study the density profile
in the open system. In addition, an analysis of the finite-size corrections
to the equilibrium measure led to an alternative formula for the diffusivity that 
can be used for high precision numerical measurements.

Nongradient lattice gases are the rule, not the exception, and the procedure used 
in this work can be extended in various directions. One could study
one-dimensional processes with local hopping rates that depend in an arbitrary way on the occupation
numbers of the original and of the target site. Further, the variational formula is valid in arbitrary
spatial dimension. For instance, one can apply it to study the diffusion coefficient and conductivity 
of kinetically constrained lattice gases \cite{KCM:rev} in two and higher dimensions.

Different variational methods have been applied to a number of lattice gas models in Refs. \cite{bib:variational-other-models}, while techniques similar to ours have been recently introduced to approximate the thermal conductivity for stochastic energy exchange models \cite{bib:GG,bib:Sasada}. It would be interesting to clarify connections between these approaches.  Another avenue for further research concerns the phenomenon of self-diffusion
describing the diffusivity of a tagged particle in a lattice gas. The coefficient of self-diffusion 
is unknown even for diffusive lattice gases satisfying the gradient condition, 
e.g. for the simple exclusion process in two and higher dimensions. The coefficient of self-diffusion 
can be written in a variational form \cite{bib:Spohn2}, and it is perhaps possible to derive 
excellent analytical approximations for this coefficient using a procedure similar to the one developed 
in this article.

\acknowledgements
CA thanks Isfahan University of Technology, Institute for Research in Fundamental Sciences (IPM), 
and IPhT, CEA Saclay for hospitality. The work of PLK was partially supported by the BSF Grant No. 2012145.
PLK is also grateful to IPhT, CEA Saclay for excellent working conditions, and to Grant Number ANR-10-LABX-0039-PALM for supporting the visits.  We are thankful to S. Mallick for a careful reading of the manuscript.

\appendix 

\section{Extreme versions of the 2-GEP}
\label{app:pecular}

For the 2-GEP, the expressions \eqref{eq:single-weights} for the one-site probabilities indicate that unusual behaviors may occur in the extreme cases of $a=0$ and $a=\infty$. The latter case, $a=\infty$, is realized when $p_{20}=0$. The 2-GEP in this case is not a diffusive lattice gas. Instead, starting from an arbitrary initial configuration, the system gets trapped in a jammed configuration like
\begin{equation*}
\ldots 002020002222022202022200002022202 \ldots
\end{equation*}
where each site is empty or occupied by two particles. In the long-time limit the configuration looks like
\begin{equation*}
\ldots 020201022221222020222001020222102 \ldots
\end{equation*}
These 1's diffuse, $10\leftrightarrow 01$ and $12\leftrightarrow 21$, 
and annihilate, $11\to 20$ or $11\to 02$, i.e., 1's disappear due to the single-species diffusion-controlled annihilation (see \cite{bib:KRB-N} for a review). In one dimension, the density of 1's decays as $t^{-1/2}$. Thus, the 2-GEP with $p_{20}=0$ is not diffusive and the system gets trapped in a jammed final state. 

The case of $a=0$ is realized when $p_{11}=0$ (see also \cite{bib:AM}).  Note that, in this case, $ D(\rho) $ is not continuous at $ \rho = 1 $, and correspondingly $ \frac{d^2 \mathcal F}{ d \rho^2 } $ diverges at this point. Elsewhere the Einstein relation \eqref{EinsteinRelation} should be satisfied.  If the global density is $\rho<1$ and additionally we start with a configuration where each site is occupied by at most one particle, $\tau_i(0)\leq 1$, such occupation arrangement will persist, so we recover the classic exclusion process. Similarly if the global density is $\rho>1$ and additionally we start with a configuration where each site is occupied by at least one particle, $\tau_i(0)\geq 1$, the process can be mapped into the classic exclusion process after interpreting a site with two particles as occupied and a site with one particle as empty. Generally when $\rho\ne 1$, the 2-GEP with $p_{11}=0$ essentially reduces to the classic exclusion process after an earlier regime when the system reaches the state with $\tau_i\leq 1$ (if $\rho<1$) or $\tau_i\geq 1$ (if $\rho>1$). 
A novel behavior can only occur when $\rho=1$. In the long time, the configuration will look like
\begin{equation*}
\ldots 11111011111111112111110111111111211111 \ldots
\end{equation*}
with 0's and 2's diffusing and annihilating upon colliding: $02\to 11$, $20\to 11$. For this two-species diffusion-controlled annihilation process (see \cite{bib:KRB-N} for a review), the density of defects (0's and 2's) is known to decay as $t^{-1/4}$. (This is certainly valid when both species of defects, 0's and 2's, have the same diffusion coefficients, that is when $p_{10}=p_{21}$.) Thus at the half-filling, $\rho=1$, the 2-GEP with $p_{11}=0$ algebraically approaches to the uniform final state where each site is occupied by one particle.

\section{Validity of the product measure for the 3-GEP}
\label{app:3-GEP}

Here we briefly consider the 3-GEP to appreciate possible qualitative differences with the 2-GEP. The 3-GEP is generally characterized by nine non-vanishing rates $ p_{rs} $ with $ r=1,2,3 $ and $s=0,1,2 $. The detailed-balance relations \eqref{DB} lead again to \eqref{eq:aX1X1-X0X2=0} and additionally to
\begin{equation}
\label{XXXX}
p_{30} W_3 W_0 = p_{12} W_1 W_2\,, \quad 
p_{31} W_3 W_1 = p_{22} W_2^2\,.
\end{equation}
The equilibrium probabilities take the form \eqref{weights}, viz. 
\begin{equation*}
 W_0 = \frac{1}{Z} \,, ~
 W_1= \frac{\lambda}{Z} \,, ~
 W_2 = a_2 \frac{ \lambda^2}{ Z} \,,~
 W_3 = a_3 \frac{ \lambda^3}{ Z} \,,
\end{equation*}
with $a_3=a_2 p_{12}/p_{30}$ and the same $a_2 = p_{11}/p_{20}$ as for the 2-GEP. All this is valid, however, only when the constraint
\begin{equation}
\label{3GEP:constraint}
 p_{31}p_{12}p_{20} = p_{30}p_{22}p_{11}
\end{equation}
following from \eqref{XXXX} is obeyed. Thus for the 3-GEP, the factorization holds when the constraint \eqref{3GEP:constraint} is satisfied; similar conditions for the $k$-GEPs with $k\ge 4$ are discussed in Ref.~\cite{bib:Cocozza-Thivent}.

\section{Details for $n=3$}
\label{sec:n=3}
 
Solving $ \frac{ \partial Q }{ \partial f_{\xi\eta\zeta} } = 0 $ ($ \xi , \eta,\zeta \in \{ 0,1,2 \} $), 
one arrives at 13 homogeneous relations (we use the shorthand notation $g_{abc}=f_{abc}+f_{cba}$)
\begin{align*}
 f_{020} + f_{202} & = 2 f_{111} = f_{000} + f_{222}, \\
g_{001} - g_{011} &= 2f_{101} - 2f_{111} = 2f_{000} - 2f_{010} , \\ 
 f_{111} - f_{212} &= f_{121} - f_{222} = g_{122} - 2 f_{222} , \\ 
 f_{111}-f_{212} & = g_{011} - g_{012} = g_{021} - g_{022}, \\ 
 g_{001} - f_{101} &= g_{002} - f_{202} = 2 f_{111}-f_{222}, \\ 
 f_{222} +f_{020} &= g_{022} , \\
 f_{101}+f_{202} &= g_{102}, \\ 
 f_{111}+f_{212} &= g_{112}, 
 \end{align*}
and 4 inhomogeneous relations 
\begin{align*}
 f_{1 0 1}-f_{1 0 0}+f_{2 0 0}-f_{2 0 1}&= U_1 , \\ 
f_{121} -f_{1 2 0} +f_{2 2 0}-f_{2 2 1}
&=U_2 , \\ 
f_{1 2 0} -f_{1 0 1}+f_{2 0 1}-f_{2 1 0}+f_{2 1 2}-f_{222}
&= U_3 , \\ 
 f_{111}-f_{1 1 0}+f_{2 1 0}-f_{2 1 1} &=U_4 \,.
\end{align*} 
Here $ U_i = A_i Z/B $, $A_i$ are polynomials of the 4th degree in $ \lambda $ and $B$ is the polynomial of the 6th degree. For the particle-uniform rates \eqref{eq:rates}, 
\begin{align*} 
A_1 &= 70+164 \lambda +175 \lambda ^2+95 \lambda ^3+24 \lambda ^4 , \\
A_2 &= 148+340 \lambda +336 \lambda ^2+164 \lambda ^3+35 \lambda ^4 ,\\
A_3 &= 350+960 \lambda +1039 \lambda ^2+527 \lambda ^3+105 \lambda ^4 , \\
A_4 &= 50+106 \lambda +97 \lambda ^2+45 \lambda ^3+10 \lambda ^4 . 
\end{align*} 
For the site-uniform rates \eqref{eq:another},
\begin{align*}
A_1&= 15+42 \lambda +64 \lambda ^2+50 \lambda ^3+21 \lambda ^4 ,\\
A_2&= 21+50 \lambda +64 \lambda ^2+42 \lambda ^3+15 \lambda ^4 ,\\
A_3&= 3 (1+\lambda )^2 \left(15+22 \lambda +15 \lambda ^2\right), \\
A_4&= -3 (1+\lambda )^4 . 
\end{align*}
For the misanthrope process $ A_i = 0 $. The expressions for $B(\lambda)$ for the particle-uniform and site-uniform rates are presented in subsection \ref{sub:3}.

\section{Microscopic coupling to the reservoirs}
\label{sec:app-coupling}

At the left boundary, the site number 1 of the open system is
connected to a reservoir at density $\rho_0$. Particles are injected to site 1 at rate $\alpha(\tau_1)$ 
and removed from site 1 at rate $\gamma(\tau_1)$. These rates depend on the occupation $\tau_1$ of site 1,
so there are four different values: $\alpha(0)$, $\alpha(1)$, $\gamma(1)$ and $\gamma(2)$. 
Imposing a local equilibrium condition with the left reservoir at density $\rho_0$ implies the following
constraints: 
\begin{equation}
\label{rates:left}
\begin{split}
\alpha(0) W_0 (\rho_0) &= \gamma(1) W_1(\rho_0),\\
\alpha(1) W_1(\rho_0) &= \gamma(2) W_2(\rho_0) . 
\end{split}
\end{equation}
The weights $ W_0$, $W_1$ and $W_2$ are the 2-GEP equilibrium probabilities \eqref{eq:single-weights} at 
density $\rho_0$, the corresponding fugacity $\lambda_0 =\lambda(\rho_0)$ is obtained 
from \eqref{eq:lambda=}. Relations \eqref{rates:left} leave some freedom in choosing the boundary rates. 
The following choice is suitable:
\begin{equation*}
\label{eq:leftboundary}
\begin{split}
 \alpha(0) &= p_{10} W_1(\rho_0) + p_{20} W_2(\rho_0) = \frac{
 p_{10} \lambda_0 + p_{11} \lambda_0^2 }{Z}\, ,\\ 
 \alpha(1) &= p_{11} W_1(\rho_0) + p_{21}
 W_2(\rho_0) = \frac{ p_{11} \lambda_0 + \frac{ p_{21}p_{11}}{p_{20}} \lambda_0^2 }{Z} \, , \\ 
 \gamma(1) &= p_{10}
 W_0(\rho_0) + p_{11} W_1(\rho_0) = \frac{ p_{10} + p_{11}
 \lambda_0}{Z} \, , \\ 
 \gamma(2) &= p_{20} W_0(\rho_0) + p_{21} W_1(\rho_0) = \frac{ p_{20} + p_{21} \lambda_0}{Z} .
\end{split}
\end{equation*}
The local detailed balance relations are readily checked: 
\begin{align*}
& \frac{\alpha(0)}{ \gamma(1)} = \lambda_0 = \frac{W_1(\rho_0)}{ W_0 (\rho_0)} \, ,\\
& \frac{\alpha(1)}{ \gamma(2)} = \lambda_0\,\frac{p_{11}}{p_{20}} = \frac{W_2(\rho_0)}{W_1(\rho_0)} \, . 
\end{align*}

At the right boundary, the system is at local equilibrium with a
reservoir at density $\rho_L$; particles are extracted from site
$L-1$ at rate $\beta(\tau_{L-1})$ and injected with rate
 $\delta(\tau_{L-1})$. A suitable choice of boundary rates is 
\begin{align} 
 \beta(r) &= \sum_{ s = 0,1} p_{ r s } W_s (\rho_L ) \quad (
 r=\tau_{L-1} ),\\ \delta(r) &= \sum_{ s =1,2 } p_{ s r } W_s (\rho_L
 ) \quad ( r=\tau_{L-1} ) , 
\label{eq:rightboundary}
\end{align} 
where now the weights $W_s$ are evaluated for the value of
$\lambda_L$, obtained from $\rho_L$ using \eqref{eq:lambda=}.

In general, the product measure does not give the correct stationary
state, except for the case $ \rho_0=\rho_L $. The difference between
the reservoir densities $ \rho_0 $ and $\rho_L $ gives a finite
current even in the stationary state. 

Simulations shown in this work were performed in the situation when
the boundary densities are extreme, viz. the left reservoir is fully
packed, $ \rho_0 = 2 $, while the right reservoir is empty, $ \rho_L
=0 $; the corresponding values of the fugacity are $ \lambda =\infty $
and $ \lambda = 0 $, respectively. Accordingly the injection and
extraction rates are 
\begin{align} 
 \alpha = p_{2s} \ (s=\tau_1),\ \beta = p_{ r 0 } \ (r=\tau_{L-1})
 ,\ \gamma = \delta = 0 . 
\end{align} 

\section{Derivation of Eqs.~(\ref{eq:Ji=hatJi+(ppp)(CC)})--(\ref{eq:hatJi=})}
\label{sec:app-currents}

First, we show that Eq.~\eqref{eq:Ji=} is equivalent to 
\begin{eqnarray}
\label{eq:gradient-part}
 J_i &=& \sum_{r=1,2}p_{r0} \mathbb P [ \tau_i = r ] -
 \sum_{r=1,2}p_{r0} \mathbb P [ \tau_{i+1} = r ] \nonumber\\ 
 &+& (p_{21}+p_{10}-p_{20}) (X^{21}_i-X^{12}_i ). 
\end{eqnarray}
Note that the top line on the right-hand side of
Eq.~\eqref{eq:gradient-part} is in a gradient form, but the bottom
line is not (unless $p_{10}-p_{20}+p_{21} = 0$). We use the following
identities
\begin{align} \label{eq:deltas}
\begin{split}
 \mathbb P [\tau_i = 0] &= \langle \delta_{\tau_i,0} \rangle =
 \frac{1}{2} \langle (1- \tau_i ) (2- \tau_i ) \rangle, \nonumber
 \\ \mathbb P [\tau_i = 1] &= \langle \delta_{\tau_i,1} \rangle =
 \langle \tau_i (2- \tau_i ) \rangle , \\ \mathbb P [\tau_i = 2] &=
 \langle \delta_{\tau_i,2} \rangle = \frac{1}{2}\langle \tau_i
 (\tau_i -1 ) \rangle , 
\end{split}
\end{align}
for $ \tau_i \in \{0,1,2\} $. The Kronecker delta symbol $
\delta_{\tau_i,r} $ indicates if site $i$ is occupied by $r$
particles. Similar expressions are readily written for two-point
probabilities $ \mathbb P [\tau_i = r \wedge \tau_{i+1} = s]$. The
expression \eqref{eq:Ji=} for the current $ J_i $ is thus given by
the expectation value
\begin{align}
J_i &= \frac{1}{2}\Big\langle ( \tau_i-\tau_{i+1}) \Big[ 2p_{10}
 (2-\tau_i-\tau_{i+1} ) \nonumber \\ &+ 
 p_{20} (\tau_i+\tau_{i+1} -1) + ( p_{21}+ p_{10} -p_{20} )\tau_i \tau_{i+1}
 \Big] \Big\rangle\,. \nonumber
\end{align}
Introducing the observables 
\begin{align}
 N_1(\tau_i,\tau_{i+1}) &= ( \tau_i-\tau_{i+1}) (2-\tau_i-\tau_{i+1} )
 \nonumber \\ &= \delta_{\tau_i,1 } - \delta_{\tau_{i+1},1 } \, ,
 \nonumber \\ N_2(\tau_i,\tau_{i+1})&= \tfrac{1}{2} (
 \tau_i-\tau_{i+1}) ( \tau_i+\tau_{i+1} -1) \nonumber \\ &=
 \delta_{\tau_i,2 } - \delta_{\tau_{i+1},2 }\, , \nonumber
 \\ N_3(\tau_i,\tau_{i+1})&= \tfrac{1}{2} ( \tau_i-\tau_{i+1})
 \tau_i \tau_{i+1}\nonumber \\ &= \delta_{\tau_i,2 }
 \delta_{\tau_{i+1},1 }- \delta_{\tau_i,1 } \delta_{\tau_{i+1},2 } \,
 , \nonumber 
\end{align}
we obtain 
\begin{eqnarray*}
 J_i &=& p_{10 } \langle N_1 ( \tau_i, \tau_{i+1}) \rangle +
 p_{20 } \langle N_2 ( \tau_i, \tau_{i+1}) \rangle \\
 &+& ( p_{21} +p_{10} -p_{20} ) \langle N_3 ( \tau_i, \tau_{i+1}) \rangle
\end{eqnarray*}
leading to \eqref{eq:gradient-part}. Performing analogous calculations for $\widehat J_i$ defined in \eqref{eq:hatJi=}, we find that $ J_i-\widehat J_i $ 
is given by 
\begin{equation}
\label{N:long}
\langle N_3 ( \tau_i , \tau_{i+1} ) \rangle - \langle \delta_{\tau_i,2 } \rangle \langle
 \delta_{\tau_{i+1},1 } \rangle + \langle \delta_{\tau_i,1 } \rangle
 \langle \delta_{\tau_{i+1},2 } \rangle
\end{equation}
times $( p_{21}+p_{10} -p_{20} )$. One can verify that \eqref{N:long} is identical to $ C_i^{21}-C_i^{12}$. This completes the proof of \eqref{eq:Ji=hatJi+(ppp)(CC)}--\eqref{eq:hatJi=}.

\vspace{5mm}
\section{Continuous limit of $ \widehat J_i $}
\label{sec:limJhat}

We now derive \eqref{grad-J}, the continuous limit 
of $ \widehat J_i $. We use the scaling form \eqref{eq:P=X+kappa/L} and obtain 
\begin{align}
& Y_i^{rs} - Y_i^{sr} \simeq W_r ( \rho_i ) W_s ( \rho_{i+1} ) - W_s ( \rho_i ) W_r ( \rho_{i+1} ) \nonumber \\
& + \frac{1}{L} \Big[ \kappa_r \Big( \frac{i}{L} \Big) \kappa_s \Big( \frac{i+1}{L} \Big) 
 - \kappa_s \Big( \frac{i}{L} \Big) \kappa_r \Big( \frac{i+1}{L} \Big) \Big] .
\label{eq:1/L[krks-kskr]}
\end{align}
The second line on the right-hand side of Eq.~\eqref{eq:1/L[krks-kskr]} vanishes as $O(L^{-2})$. To simplify the first line in Eq.~\eqref{eq:1/L[krks-kskr]} we write the Taylor expansion 
\begin{align}
 W_\ell ( \rho_{i+1} ) \simeq W_\ell ( \rho ) + \frac{1}{L} \frac{d\rho}{dx}\frac{dW_\ell}{d\rho}
 \quad (\ell=r,s). 
\end{align}
Using this result and keeping only $O(L^{-1})$ terms we simplify Eq.~\eqref{eq:1/L[krks-kskr]} and arrive at 
\begin{align}
\label{hat:I}
 \widehat{J}(x) \simeq  \frac{1}{L} \frac{d\rho}{dx} \sum_{ r =1,2 \atop s=0,1} p_{rs} 
 \left[ W_r \frac{dW_s}{d\rho} - W_s \frac{dW_r}{d\rho} \right]. 
\end{align}
We emphasize that the finite-size corrections $ \kappa_r $ cancel out in \eqref{hat:I}. 
Substituting \eqref{eq:single-weights} into \eqref{hat:I} we obtain 
\begin{equation*}
 \widehat{J}(x) \simeq -  \frac{1}{L} \frac{d\rho}{dx} \frac{d\lambda }{d \rho} \frac{ 1 }{ \lambda }
 \sum_{ 1\le r \le2 \atop 0\le s \le 1 } p_{rs} W_r W_s 
\end{equation*}
which indeed reduces to \eqref{grad-J} once we recall that $ D_0 $ is given by \eqref{eq:D0=}.


\begin{thebibliography}{00}
\bibitem{prod_measure} 
 By ``product measure'' we mean that in the equilibrium correlation functions factorize, e.g. the pair correlation
 function satisfies $\langle \tau_i \tau_j\rangle = \langle \tau_i \rangle \langle \tau_j \rangle$ for $i\ne j$.
\end{thebibliography}
\end{document}